\documentclass[11pt]{article}

\usepackage{preamble}

\title{The CNF Encoding Complexity of Boolean Functions}
\date{}

\author{Andrew Krapivin\thanks{\url{andrew@krapivin.net}}~\;\orcidlink{0009-0003-0227-7660}}
\author{Benjamin Przybocki\thanks{\url{benjamin.przybocki@gmail.com}}~\;\orcidlink{0009-0007-5489-1733}}
\author{Bernardo Subercaseaux\thanks{\url{bersub@cmu.edu}}~\;\orcidlink{0000-0003-2295-1299}}

\affil{Carnegie Mellon University}

\date{}

\newcommand{\lrp}[1]{\left( #1 \right)}
\DeclareMathOperator{\abip}{\mathnormal{\alpha}_{\textsf{b}}}

\begin{document}
\maketitle
\begin{abstract}
    We study the minimum number of clauses required to encode a boolean function into a CNF formula with auxiliary variables.
    Our main result is that if a boolean function can be evaluated in nondeterministic time $t$ (in the RAM model) with $m$ bits of advice such that at most $b$ bits of advice are accessed for each input, then it admits a CNF encoding with $\widetilde{O}(t + \sqrt{mb})$ clauses. This constitutes a nonuniform generalization of Robson's (TCS'91) variant of the Cook--Levin theorem, for which a na\"ive  generalization would yield $\widetilde{O}(t+m)$ clauses; our applications leverage $b = o(m)$ to obtain significantly better bounds. 
    Since preprocessed data structures can be modeled as computation with advice, our result allows us to translate space- and time-efficient data structures to compact CNF encodings.
     As an application, we derandomize a data structure of Larsen and Williams (SODA'17) to show that every $k$-CNF function on $n$ inputs can be encoded with $n^k/2^{\Omega_k(\sqrt{\lg n})}$ clauses. Notably, despite encoding a $k$-CNF function, the encoding itself necessarily contains clauses of unbounded width, since an information-theoretic argument shows that the minimum number of bounded-width clauses to encode such functions is $\Omega_k(n^{k}/\lg n)$ in the worst case.  In general terms, our results imply that for a wide variety of NP-complete problems, one can reduce to~\textsf{CNF-SAT} in a way that results in asymptotically fewer clauses than previously known methods.
\end{abstract}

\thispagestyle{empty}
\setcounter{page}{0}

\setcounter{tocdepth}{2}
\thispagestyle{empty}
\setcounter{page}{0}
\newpage

\section{Introduction}\label{sec:intro}
Given a boolean function $f$, how compactly can we ``encode'' $f$ as a CNF formula (see~\Cref{def:encoding})? A partial answer was provided first by the Cook--Levin theorem~\cite{Cook71,Levin1973}, which implies that if $f$ can be computed nondeterministically in $n$ steps, then it can be encoded as a CNF instance of size $n^{O(1)}$. 
Since then, more compact encodings of size $\widetilde{O}(n)$ have been devised and extended to different models of computation~\cite{schnorr1978satisfiability,pippenger1979relations,cook1988short,robson1991ot}. These optimized Cook--Levin theorems have several complexity-theoretic applications, including showing that \textsf{CNF-SAT} is complete for nondeterministic quasilinear time~\cite{schnorr1978satisfiability,nearly-linear} and proving time-space tradeoffs for \textsf{CNF-SAT}~\cite{fortnow}. The problem of compactly encoding boolean functions has also acquired tremendous practical utility with the advent of SAT solvers, where efficient encodings are often make-or-break for whether a problem can be solved in practice~\cite{subercaseauxPackingChromaticNumber2023,kirchweger2025finiteinfinitesharperasymptotic,happyEnding,unfolding,cliquewidth}; empirically, smaller encodings tend to perform better in practice~\cite{bjorkSuccessfulSATEncoding2011a,abioParametricApproachSmaller2013c}.\footnote{The claim here is \emph{not} that formulas with fewer clauses are easier to solve---random formulas with hundreds of clauses tend to be much harder than structured formulas from industry with millions of clauses---but rather that for a \emph{fixed} problem, performance tends to improve by choosing a smaller encoding~\cite{Prestwich21}.}

Consequently, the SAT community has studied encodings for a variety of concrete functions, including both classical NP-complete problems  (e.g., Hamiltonian cycle~\cite{Zhou20,Zhou24}, graph coloring~\cite{FaberJM24}, $k$-clique~\cite{IgnatievMM17,subercaseaux2025asymptoticallysmallerencodingsgraph}, and many others~\cite{CourcelleD26}) as well as elementary constraints such as parity~\cite{EmdinKMS22} or \emph{at-most-$k$}~\cite{sinz,kucera,NearOptimalCardinalitySubmitted}, which serve as building blocks of larger encodings. However, despite the theoretical and practical importance of CNF encodings, this model of computation has yet to receive a systematic theoretical treatment that focuses on large classes of boolean functions rather than particular instances, which is in stark contrast to other models of computation like boolean circuits~\cite{boolean-complexity}. In this paper, we prove fundamental results about the complexity of boolean functions in this model and show how this complexity metric relates to complexities in other nonuniform models. We hope to convince the reader that the theory of CNF encodings is much deeper than has previously been appreciated. In particular, we show that the optimized Cook--Levin results mentioned above are not the end of the story.

A CNF \emph{encoding} of a boolean function $f$ is a CNF formula with nondeterministic variables that computes $f$. Formally:
\begin{definition}[Encoding]\label{def:encoding}
    Given a boolean function $f \colon \{\zero,\one\}^n \to \{\zero, \one\}$, we say that a triple $(\varphi, X, Y)$ is an \emph{encoding} of $f$ if:
    (i) $X := (x_1, \dots, x_n)$ is a sequence of propositional variables, and $Y$ is a set of propositional variables disjoint from $X$;
     (ii) $\varphi$ is a CNF formula over variables $X \sqcup Y$; and (iii) for every assignment $\tau \colon X \to \{\zero, \one\}$,
        \[
        f(\tau(x_1), \dots, \tau(x_n)) = \one \iff  \exists \theta \colon Y \to \{\zero, \one\}, \; (\tau \cup \theta) \models \varphi.
        \]
    The variables in $X$ are called \emph{input} variables, and those in $Y$ are called \emph{auxiliary} variables.
\end{definition}
Thus, CNF encodings constitute a nondeterministic, nonuniform model of computation. We say that the \emph{CNF encoding complexity} of a boolean function $f$, denoted $\enc(f)$, is the minimum number of clauses in a CNF encoding of $f$. In \Cref{sec-why}, we elaborate on why we choose to focus on the clause count instead of, say, the total number of literals in a CNF encoding.

One of the few theoretical results known about encoding complexity is due to Tseitin~\cite{tseytin}: if the (nondeterministic) circuit complexity of $f$ is $n$, then $\enc(f) = O(n)$. Each of our main results implies that the converse of Tseitin's theorem fails strongly: there are boolean functions for which $\enc(f)$ is much smaller than its (nondeterministic) circuit complexity. In \Cref{sec-characterization}, we characterize $\enc(f)$ exactly in terms of nondeterministic \emph{conjunctive} circuit complexity: whenever $\enc(f)\ge2$, it equals the minimum number of wires entering $\land$-gates in a nondeterministic De Morgan circuit computing $f$. Thus, all of our results also have circuit complexity analogues.

\subsection{Our results}

Our first result relates encoding complexity to the time complexity of nondeterministic RAM machines with advice. Robson~\cite{robson1991ot} showed the following variant of the Cook--Levin theorem: if $f\colon \{\zero, \one\}^n \to \{\zero, \one\}$ can be computed by a nondeterministic RAM machine in time $t \ge n$, then $\enc(f) = O(t \lg t)$, and such an encoding can be constructed in polynomial time (strictly speaking, if the machine has program size $p$, Robson yields $O((t+p) \lg(t+p))$). However, there is a precise sense in which Robson's result does not capture the full power of CNF encodings; nondeterministic RAM is a uniform model of computation, whereas CNF encodings constitute a nonuniform model. Thus, it is natural to instead translate nondeterministic RAM machines \emph{with advice} (a model we call \emph{ANRAM}) to CNF encodings. A na\"ive extension of Robson's result would state that if $f$ can be computed by an ANRAM machine in time $t$ with $m$ bits of advice, then $\enc(f) = \widetilde{O}(t + m)$.\footnote{Throughout this paper, we use $\widetilde{O}(g)$ to denote $O(g\cdot \mathrm{polylog}(g))$ for any function $g$, and $f = \widetilde{\Omega}(g)$ meaning $g = \widetilde{{O}}(f)$.}  We show the following quantitative improvement:

\begin{restatable}{theorem}{robsonadvice} \label{thm:robson-advice}
    Let $f\colon \{\zero, \one\}^n \to \{\zero, \one\}$, and suppose that $f$ can be evaluated in time $t \geq n$ with $m$ bits of advice by an ANRAM machine that accesses at most $b$ bits of the advice for each input. Then, $\enc(f) = \widetilde{O}(t + \sqrt{mb})$. Furthermore, such an encoding can be constructed deterministically in time polynomial in $n+t+m$ given the advice string.
\end{restatable}

We find~\Cref{thm:robson-advice} remarkable. It provides a tight bound on the number of clauses (up to polylogarithmic factors) for a wide variety of functions (technically speaking, even for a $1-o(1)$ fraction of all boolean functions), and yields encodings with fewer clauses for a variety of classical $\mathsf{NP}$-complete problems (we discuss some in~\Cref{sec-applications}). In contrast, applying Robson's result yields quadratically worse bounds for almost all functions. 
Indeed, suppose one takes $f\colon \{\zero, \one\}^n \to \{\zero, \one\}$ to be an arbitrary boolean function. Then, letting the advice string $A$ be the string of length $2^n$ encoding $f$'s truth table, we can compute $f$ by reading the input $\vec{x}$ and consulting $A[\vec{x}] = f(\vec{x})$; this way $t = O(n), m = 2^n,$ and $b = 1$, which yields $\enc(f) = \widetilde{O}(\sqrt{2^n})$. In contrast, an application of Robson's theorem would yield the trivial bound $\enc(f) = \widetilde{O}(2^n)$. \footnote{Indeed, there are $2^{2^n}$ boolean functions on $n$ variables, and each of them requires a different NRAM program, so the majority of them require program size $p = \widetilde{\Omega}(2^n)$.}

In fact, a simple lower bound shows that~\Cref{thm:robson-advice} is not too far from optimal, meaning roughly that for every boolean function $f$, there is some ANRAM machine for which applying \Cref{thm:robson-advice} yields an encoding of size at most $\enc(f)^{3/2}$.

\begin{proposition} 
   Let  $f\colon \{\zero, \one\}^n \to \{\zero, \one\}$ be a boolean function with $\enc(f) = \Omega(n)$, let $(M^\star_f, A^\star_f)$ be an ANRAM machine that computes $f$ and minimizes $t+\sqrt{mb}$ over all possible (machine, advice) pairs, and let $t^\star, m^\star, b^\star$ be the corresponding optimal parameters. Then, 
    \[ 
        \enc(f) = \widetilde{\Omega}\left(\left(t^\star + \sqrt{m^\star b^\star}\right)^{2/3}\right).
    \]
\end{proposition}
\begin{proof}
    Let $F$ be a formula that encodes $f$ with $\enc(f)$ clauses, and $Y$ be its set of auxiliary variables. We can further assume that $|Y| \leq \enc(f)$; we prove that this assumption is without loss of generality in~\Cref{prop:auxiliaries}.
    Then, consider the ANRAM machine receiving $F$ as advice, and upon receiving an input $\vec{x}$, proceed as follows. First, guess an assignment $\theta$ for each of the auxiliary variables in time $O(|Y|) = O(\enc(f))$. Next, 
    for each clause in $F$, nondeterministically guess which of its at most $n+\enc(f)$ literals satisfies it; guessing this index takes time $O(\lg(\enc(f)))$. Then, check that the polarity of that literal matches its value on the input, or on $\theta$ if it corresponds to an auxiliary variable, which can be done in time $O(\lg(\enc(f)))$, and accept if and only if all such checks succeed.  Thus, the total time is $t = O(\enc(f)\lg(\enc(f)))$, the advice length is $m = \widetilde{O}(\enc(f)^2)$ and the number of bits read is $b = \widetilde{O}(\enc(f))$. We thus have  
    \[ 
        t + \sqrt{mb} = \widetilde{O}(\enc(f)) + \sqrt{\widetilde{O}(\enc(f)^2) \cdot \widetilde{O}(\enc(f))} = \widetilde{O}(\enc(f)^{3/2}).
    \] 
    Since $t + \sqrt{mb} \geq t^\star + \sqrt{m^\star b^\star}$ by definition, we obtain $\widetilde{O}(\enc(f)^{3/2}) \geq t^\star + \sqrt{m^\star b^\star}$ from which the result follows.
\end{proof}

More generally, \Cref{thm:robson-advice} has a natural interpretation in terms of data structures. Suppose we have an algorithm to compute $f$ that has access to a data structure $D_f$ which uses $m$ bits of memory, and upon receiving an input $\vec{x}$, nondeterministically computes $f(\vec{x})$ in time $t$ by querying at most $b$ bits of $D_f$.
Then, \Cref{thm:robson-advice} says that $f$ must have a small CNF encoding, regardless of how long it takes to build $D_f$. As a more interesting application, consider the functions $\mathsf{IS}_G \colon \{\zero,\one\}^n \to \{\zero, \one\}$, each parametrized by some graph $G$ on $n$ vertices $v_1, \ldots, v_n$, defined by
\[
    \mathsf{IS}_G(x_1, \dots, x_n) = \one \iff \{ v_i \mid x_i = \one \} \text{ is an independent set of } G.
\]
Larsen and Williams~\cite[Thm. 3.2]{larsenWilliams} showed that there is a randomized data structure using $O(n^2)$ bits of space that computes this function in time $\frac{n^2}{2^{\Omega(\sqrt{\lg n})}}$. We show that this data structure can be derandomized, and therefore \Cref{thm:robson-advice} implies that $\enc(\mathsf{IS}_G) = \frac{n^2}{2^{\Omega(\sqrt{\lg n})}}$. It is not hard to see that this implies that every 2-CNF function has a CNF encoding of size $\frac{n^2}{2^{\Omega(\sqrt{\lg n})}}$. More generally, we show the following:
\begin{restatable}{theorem}{thmquasi}\label{thm-quasi}
    If $f \colon \{\zero,\one\}^n \to \{\zero,\one\}$ is a $k$-CNF function for some $k \ge 2$, then $\enc(f) = \frac{n^k}{2^{\Omega_k(\sqrt{\lg n})}}$.
\end{restatable}

The class of $k$-CNF functions (i.e., representable by CNF formulas where each clause has width exactly $k$) is not only theoretically interesting~\cite{unate} but is also of great practical importance in SAT solving, particularly for $k=2$. This is because the encodings for many NP-complete graph problems (e.g., vertex cover, $k$-coloring, $k$-clique, etc.) are dominated by a 2-CNF component~\cite{subercaseaux2025asymptoticallysmallerencodingsgraph}. \Cref{thm-quasi} implies that, for a fixed graph, all of these problems have CNF encodings with $\frac{n^2}{2^{\Omega_k(\sqrt{\lg n})}}$ clauses. Previously, the best known bound was $O(n^2 / \lg n)$~\cite{subercaseaux2025asymptoticallysmallerencodingsgraph,PreprocessingSubmitted}, and more generally, hypergraph decompositions give encodings for $k$-CNF functions with $O_k(n^k/\lg n)$ clauses~\cite{ep-hypergraph,stoc}. Interestingly, a counting argument shows that if we restrict ourselves to encodings that are themselves $k$-CNF (or even bounded width), then some $k$-CNF functions require $\Omega_k(n^k / \lg n)$ clauses. Thus, \Cref{thm-quasi} rather surprisingly implies that clauses of unbounded width are helpful for encoding $k$-CNF functions.

Larsen and Williams' data structure requires superpolynomial preprocessing time, so we do not know how to construct the encoding given by \Cref{thm-quasi} in polynomial time. However, we do have a weaker bound, albeit still improving on the $\Theta_k(n^k / \lg n)$ bound for bounded-width encodings, for which we have a polynomial-time construction:
\begin{restatable}{theorem}{thmpolytime}\label{thm-polytime}
    If $f \colon \{\zero,\one\}^n \to \{\zero,\one\}$ is a $k$-CNF function for some fixed $k \ge 2$, then one can construct in deterministic polynomial time a CNF encoding of $f$ with
    \(
        O_k\left(\frac{n^k(\lg\lg n)^3}{\lg^2 n}\right)
    \)
    clauses.
\end{restatable}

The best lower bound we have for the encoding complexity of $k$-CNF functions is $\Omega_k(n^{k/2})$ clauses obtained via a counting argument (\Cref{prop-kcnf-unbounded-width}). Thus, there is a wide gap between our upper and lower bounds for this problem, and narrowing this gap is an interesting open problem.

En route to proving \Cref{thm:robson-advice}, we prove a result about encoding arbitrary boolean functions that are sparsely supported. We later sharpen this bound to the following, which may be of independent interest:
\begin{restatable}{theorem}{thmsparse} \label{thm:sparse}
    For all $1\le M\le2^{n-1}$, we have
    \[
       \enc(n,M):= \max\left\{\enc(f):f\colon\{\zero,\one\}^n\to\{\zero,\one\},\ |f^{-1}(\one)|=M\right\} = \Theta\left(n+\sqrt{M(n-\lg M )}\right).
    \]
\end{restatable}
As a direct corollary, we get an upper bound on arbitrary functions. Let $f \colon \{\zero,\one\}^n \to \{\zero,\one\}$, and consider the function $g_f \colon \{\zero,\one\}^{n+1} \to \{\zero,\one\}$ defined by $g_f(\vec{x}, y) = f(\vec{x}) \land y$, for which applying~\Cref{thm:sparse} gives $\enc(g_f) = O(\sqrt{2^n})$. It is then easy to see that any encoding for $g_f$ is also an encoding for $f$ taking $y$ as an auxiliary variable.
Thus, $\enc(f) = O(\sqrt{2^n})$, and this bound is optimal; this is a slight strengthening of the $\widetilde{O}(\sqrt{2^n})$ bound we obtained from \Cref{thm:robson-advice} above. Similar bounds are known for depth-3 unbounded fan-in circuits from~\cite{ac0} and nondeterministic branching programs~\cite{nechiporuk1962zero} (see~\Cref{sec-related}).
In contrast, applying the Tseitin transformation to an optimal (fan-in-$2$) 
circuit for $f$~\cite{Lupanov1958} only gives $\enc(f) = O(2^n/n)$. In this sense, encodings, whose clauses can be arbitrarily wide, are more directly comparable to circuits with unbounded fan-in gates or nondeterministic branching programs.

\subsection{Further applications of \Cref{thm:robson-advice}}\label{sec-applications}

To demonstrate how generally applicable \Cref{thm:robson-advice} is, we derive a few corollaries for problems of interest to the SAT community. 

\begin{corollary}\label{cor:hc}
    Let $G$ be a graph with $V(G) = [n]$, and let $\mathsf{HC}_G$ be the boolean function that receives vertices $v_1, \dots, v_n \in [n]$, each written in binary using $\lceil \lg n \rceil$ input variables, and outputs $\one$ if and only if $v_1, \dots, v_n$ are distinct and $\{v_i, v_{i+1}\} \in E(G)$ for every $i \in [n]$, where $v_{n+1} := v_1$. Then, $\enc(\mathsf{HC}_G) = \widetilde{O}(\sqrt{n|E(G)|})$.
\end{corollary}
\begin{proof}
    We may assume that $|E(G)| \geq n$, as otherwise $\mathsf{HC}_G$ is constantly $\zero$. Let the advice be a list $A$ of the edges of $G$, each stored as a pair of integers written in binary, so $m = O(|E(G)| \lg n)$. An ANRAM machine can check that $v_1, \dots, v_n$ are distinct by scanning over the vertices and maintaining an array of which have been seen so far; then, for each $i \in [n]$, it can guess an index $p$ and verify that $A[p] = \{v_i, v_{i+1}\}$ by querying $O(\lg n)$ bits of advice. Hence, $t = \widetilde{O}(n)$ and $b = O(n \lg n)$, and \Cref{thm:robson-advice} gives $\enc(\mathsf{HC}_G) = \widetilde{O}(\sqrt{n|E(G)|})$.
\end{proof}
In contrast, the most compact CNF encodings from Zhou~\cite{Zhou20} for Hamiltonian cycle use $\Theta(|E(G)| \lg n)$ clauses. We note however that Zhou uses a different representation for the input variables: instead of a binary representation with $\Theta(n \lg n)$ variables, he uses a one-hot representation with $\Theta(|E|)$ variables. Nevertheless, our encoding provides a reduction of the Hamiltonian cycle problem to SAT with fewer clauses.

\begin{corollary}\label{cor:clique}
    Let $G$ be a graph with $V(G) = [n]$ and let $k \geq 2$. Let $\mathsf{CLIQUE}_{G,k}$ be the boolean function that receives vertices $v_1, \dots, v_k \in [n]$ in binary and outputs $\one$ if and only if $\{v_i, v_j\} \in E(G)$ for all $i \neq j$. Then, $\enc(\mathsf{CLIQUE}_{G,k}) = \widetilde{O}(k\sqrt{|E(G)|}\lg n)$.
\end{corollary}
\begin{proof}
    We may assume that $|E(G)| \geq \binom{k}{2}$, as otherwise $\mathsf{CLIQUE}_{G,k}$ is constantly $\zero$. With the same advice $A$ as in \Cref{cor:hc}, an ANRAM machine can guess, for each pair $i \neq j$, an index $p$ and verify that $A[p] = \{v_i, v_j\}$. Hence, $t = O(k^2 \lg n)$ and $b = O(k^2 \lg n)$, and \Cref{thm:robson-advice} gives $\enc(\mathsf{CLIQUE}_{G,k}) = \widetilde{O}(k\sqrt{|E(G)|}\lg n)$.
\end{proof}
For comparison, a direct encoding of $k$-clique uses $\Omega(|E(G)|)$ clauses, and Subercaseaux~\cite{subercaseaux2025asymptoticallysmallerencodingsgraph} showed one can obtain $O(n^2/ \lg n)$ clauses. \Cref{cor:clique} is better than these bounds whenever $k$ is much smaller than $\sqrt{|E(G)|}$.

Our third corollary shows that randomly chosen $k$-CNF functions have much smaller encodings than the worst-case bound of \Cref{thm-quasi}, leveraging the celebrated result of Balogh, Dong, Lidick\'y, Mani, and Zhao~\cite{unate} who showed that almost all $k$-CNF functions are \emph{unate}, i.e., antitone after negating some of the input variables.

\begin{corollary}\label{cor:random-kcnf}
    For every fixed $k \geq 2$, a uniformly random $k$-CNF function $f \colon \{\zero,\one\}^n \to \{\zero,\one\}$ satisfies $\enc(f) = \widetilde{O}(n^{k/2})$ with probability $1-o(1)$.
\end{corollary}
\begin{proof}
    By~\cite[Theorem 1.1]{unate}, the number of $k$-CNF functions on $n$ variables is $(1+o(1))2^{\binom{n}{k}+ n}$.  Thus, it suffices to show that there are $(1- o(1))2^{\binom{n}{k}+n}$ $k$-CNF functions $f$ that satisfy the desired bound. Indeed, consider the set $\mathcal{H}$ of functions of the form $\bigwedge_{i=1}^m \bigvee_{j \in S_i} \overline{x_j}$ for sets $S_1, \ldots, S_m \in \binom{[n]}{k}$, and note that $|\mathcal{H}| = 2^{\binom{n}{k}}$. 
     It is a standard fact that with high probability a uniformly random function from $\mathcal{H}$ does not accept any input with more than $s := \lceil 3 \lg n \rceil$ ones. Indeed, the expected number of satisfying inputs to a uniformly random function from $\mathcal{H}$ with Hamming weight $s$ is 
    \[
          \binom{n}{s} 2^{-\binom{s}{k}} \leq n^s 2^{-\binom{s}{2}} \leq 2^{s \lg n - s(s-1)/2} \leq 2^{-s} = o(1).
    \]
    Therefore, there is a subset $\mathcal{H}' \subseteq \mathcal{H}$ of size $(1-o(1))2^{\binom{n}{k}}$ for which no satisfying assignment has more than $s$ ones. Moreover, the subset $\mathcal{H}'' \subseteq \mathcal{H'}$ of functions for which all $n$ variables appear in their corresponding formulas satisfies $|\mathcal{H}''| = (1-o(1))|\mathcal{H}'|$.  
    Now, note that relabeling any subset of the variables of a function by their negations never changes its encoding complexity, and thus, for each function in $\mathcal{H}''$ there are $2^n$  distinct $k$-CNF functions with the same encoding complexity, and these $2^n|\mathcal{H}''|$ functions are all distinct. Therefore, it remains to see that $\enc(f) = \widetilde{O}(n^{k/2})$ for every $f \in \mathcal{H}''$.

    We show this by \Cref{thm:robson-advice}. Let the advice be the characteristic vector of the collection $S_1, \ldots, S_m$ over all $\binom{n}{k}$ size-$k$ subsets, so $m = \binom{n}{k}$.
    An ANRAM machine can deterministically scan the input and reject if it contains more than $s$ ones; otherwise, it lists the at most $s$ variables set to $\one$ and queries one advice bit for each of the at most $\binom{s}{k}$ sets of $k$ of them, accepting if and only if none of these sets has a $\one$ in the advice string. Hence, $t = \widetilde{O}(n)$ and $b = O(\lg^k n)$, and \Cref{thm:robson-advice} gives $\enc(f) = \widetilde{O}(n + \sqrt{n^k \lg^k n}) = \widetilde{O}(n^{k/2})$.
\end{proof}

We show in \Cref{prop-kcnf-unbounded-width} that, conversely, a random $k$-CNF function satisfies $\enc(f) = \Omega_k(n^{k/2})$ with high probability, so \Cref{cor:random-kcnf} is tight up to polylogarithmic factors. Curiously, the situation for worst-case $k$-CNF functions is much more complicated: there, the best bounds we know are the same $\Omega_k(n^{k/2})$ lower bound and the $\frac{n^k}{2^{\Omega_k(\sqrt{\lg n})}}$ upper bound of \Cref{thm-quasi}. Closing this gap remains an interesting open problem, and is related to the gap between the worst-case results of Larsen and Williams~\cite{larsenWilliams} for answering independent set queries in graphs, and the average case, which is much easier.

\subsection{Why study clause minimization?} \label{sec-why}

The choice of CNF encoding is known to have a dramatic impact on the performance of modern SAT solvers, allowing, for instance, the solution of long-standing problems in discrete mathematics~\cite{happyEnding,subercaseauxPackingChromaticNumber2023}. 
A major challenge, however, is to understand what properties of encodings are best correlated with empirical runtimes~\cite{bjorkSuccessfulSATEncoding2011a,ganeshUnreasonableEffectivenessSAT2021}. In 2009, Bj\"ork published interviews with several experts on practical CNF encodings, and the consensus was that clause width (which impacts the total number of literals in an encoding) was not an important factor, but the total number of clauses was more relevant.\footnote{A plausible explanation is that SAT solvers (both DPLL- and CDCL-based~\cite{Katebi2011}) spend the majority of their runtime in~\emph{unit propagation}~\cite{knuth2022taocp4b}, and for a fixed number of variables, the complexity of doing unit propagation depends heavily on the number of clauses. Interestingly, the complexity of unit propagation used to depend strongly on clause width until the development of the \emph{two watched literals} data structure~\cite{Chaff,knuth2022taocp4b}, a key ingredient of modern solvers that stores only two special literals per clause, which are sufficient to perform unit propagation, and thus reduce the impact of width.}
The encodings we present in this paper reduce the number of clauses by increasing the width, which can often be a favorable trade-off. We recently showed that CNF encodings for cardinality constraints (i.e., $\sum_{i} x_i \leq k$) with fewer but wider clauses can outperform classical encodings in some settings~\cite{NearOptimalCardinalitySubmitted}.

Furthermore, measuring complexity by the number of clauses has connections to other theoretically interesting models of computation. For example, we show in \Cref{sec-characterization} that $\enc(f)$ is equal to the minimum number of $\land$-wires (when $\enc(f)\ge2$), which is within a factor of two of the minimum number of fan-in-two $\land$-gates in a nondeterministic De Morgan circuit computing $f$. It is therefore no coincidence that our recent work on CNF encodings of cardinality constraints has yielded insights that have led to improved circuit complexity bounds~\cite{NearOptimalCardinalitySubmitted}.
Similarly, encoding complexity is related to the minimum number of constraints in integer linear programs (ILPs), since each clause $\bigvee_{i} \ell_i$ can be written as $\sum_{i} \ell_i \geq 1$. Thus, our encodings directly translate to 0/1-ILP formulations with few constraints, using auxiliary variables (cf. extension complexity~\cite{YANNAKAKIS1991441,fiorini}). 

\subsection{Related work}\label{sec-related}
The key ingredients for \Cref{thm-quasi,thm-polytime} are data structures for computing independent-set queries, which is a well-studied problem in fine-grained complexity. Williams~\cite{Williams07MatrixVectorSubquadratic} showed that there is a data structure for this task with deterministic polynomial preprocessing and $O(n^2/\lg^2 n)$ query time on a pointer machine; Blelloch, Vassilevska, and Williams~\cite{BlellochVW08CombinatorialSparseGraph} extended this work to support queries on sparse graphs. Bansal and Williams~\cite{BansalW09RegularityIndependentSet} gave faster batched queries using graph decompositions based on weak regularity. These query-time bounds do not directly give the same clause bounds in our setting, because the memory needed to represent lookup tables must also be encoded. As mentioned above, \Cref{thm-quasi} is a consequence of derandomizing a data structure of Larsen and Williams~\cite{larsenWilliams} for computing independent set queries (see \Cref{thm:derandomized}). Our proof of \Cref{thm-polytime} can be seen as yet another data structure for independent set queries, albeit implemented directly as a CNF encoding.

Our results are closely related to the extension complexity of polytopes that have a prescribed intersection with the boolean hypercube, as studied by~Hrube\v{s} and Talebanfard~\cite{Hrube2022}. They obtain an $O(\sqrt{2^n})$ extension complexity for polytopes intersecting an arbitrary $A \subseteq \{\zero, \one\}^n$, which is similar to a special case of our \Cref{thm:sparse}, although in their case the lower bound is not known to be tight (see below for other bounds of this order in related models). Furthermore, the \emph{edge polytope} of graphs, which is tightly related to the boolean function $\mathsf{IS}_G$, was studied by Fiorini, Kaibel, Pashkovich, and Theis~\cite{FIORINI201367} using biclique partitions, one of the components of our proof of \Cref{thm-polytime}.
More generally, the problem of compactly encoding $\mathsf{IS}_G$ fits into a broader landscape of \emph{graph complexity} measures~\cite{complexityGraphs,boolean-complexity}. Interestingly, most of these prior results also rely on biclique partitions, which by themselves cannot yield encodings of size $o(n^2/\lg n)$; more specifically related to our work, Subercaseaux~\cite{subercaseaux2025asymptoticallysmallerencodingsgraph} observed that biclique partitions give $\enc(\mathsf{IS}_G) = O(n^2/\lg n)$, and recently a refined analysis using rectifier networks from Przybocki, Subercaseaux, and Heule~\cite{PreprocessingSubmitted} showed $\enc(\mathsf{IS}_G) \leq (\frac{1}{4}+o(1))n^2/\lg n$, and that this cannot be improved via rectifier-network encodings.

The dense case of \Cref{thm:sparse}, namely that every boolean function on $n$ inputs has $\enc(f) = \Theta(\sqrt{2^n})$ in the worst case, has classical circuit complexity analogues: worst-case bounds of order $\sqrt{2^n}$ are well known in models of computation where gates have unbounded fan-in or some operations come for free, and in several of them the exact constant is known; see~\cite[Sec.~1.4.4]{boolean-complexity} for an overview. For instance, every boolean function on $n$ inputs can be computed by a depth-3 unbounded fan-in circuit with $(2+o(1))\sqrt{2^n}$ gates, and this is tight~\cite{ac0,sergeev-depth3}.\footnote{A na\"ive translation of Dan\v{c}\'{\i}k's unbounded fan-in circuit would yield an encoding of size $\Theta(n\sqrt{2^n})$: its first layer consists of $\sqrt{2^n}$ $\land$-gates of fan-in $n/2$, followed by a layer of $\lor$-gates and a single output $\land$-gate, and while the $\lor$-gates are free by \Cref{thm-enc-characterization}, every wire into the first-layer $\land$-gates costs a clause.} 
Nechiporuk~\cite{nechiporuk1962zero} proved that nondeterministic branching programs with 
 $(2+o(1))\sqrt{2^n}$ edges suffice for all $n$-bit functions.\footnote{The latter upper bound translates directly to our setting: replacing each node of a nondeterministic branching program by an $\lor$-gate and each literal-labeled edge by an $\land$-gate of fan-in $2$, whose inputs are the literal and the gate of the edge's tail, gives a nondeterministic De Morgan circuit whose $\land$-gates have total fan-in twice the size of the program, and hence every $f\colon\{\zero,\one\}^n\to\{\zero,\one\}$ has $\enc(f)\leq(4+o(1))\sqrt{2^n}$ by \Cref{thm-enc-characterization}. Our \Cref{prop:dense} improves the constant to $2\sqrt{2}$ and gives a lower bound of $(1/\sqrt{\lg 3}-o(1))\sqrt{2^n}$; determining the exact constant remains open.}

Recently, Sheng and Xiao have shown a surprising role of CNF encodings in the theory of kernelization, using them as ingredients to exhibit the first polynomial-sized kernel for planar directed feedback vertex set~\cite{sheng2026polynomialkernelplanardirected}, as well as a near-linear element kernel for $d$-hitting set~\cite{sheng2026nearlinearelementkerneldhitting}.

\section{Technical Overview}

\subsection{Nondeterministic RAM with advice}

In \Cref{sec:non-uniform}, we prove \Cref{thm:robson-advice}, which states that if $f \colon \{\zero,\one\}^n \to \{\zero,\one\}$ can be evaluated in time $t$ with $m$ bits of advice by an ANRAM machine that accesses at most $b$ bits of the advice for each input, then $\enc(f) = \widetilde{O}(t + \sqrt{mb})$. Our starting point is the case without advice proven by Robson~\cite{robson1991ot}, which states that $\enc(f) = \widetilde{O}(t)$ when $m=0$. Thus, the central challenge is how to encode the advice queries using only $\widetilde{O}(\sqrt{mb})$ clauses. Since the advice string can be thought of as a function $h\colon \{\zero, \one\}^{\lg m} \to \{\zero, \one\}$, and such functions can be encoded with $O(\sqrt{2^{\lg m}}) = O(\sqrt{m})$ clauses according to~\Cref{thm:sparse}, a na\"ive approach would use a total of $O(b \sqrt{m})$ clauses to query $b$ bits. We improve upon this in~\Cref{thm:checkadvice}, which provides a ``mass-production theorem'' (cf.~\cite{Ulig1974,Kretschmer23}) that roughly states we can encode $b$ queries to a function only paying a $\sqrt{b}$ multiplicative factor.
More precisely, for any boolean function $g\colon \{\zero,\one\}^k \to \{\zero,\one\}$, let $g^{\land b} \colon \{\zero,\one\}^{kb} \to \{\zero,\one\}$ be the function defined by 
\(
g^{\land b}(\vec{x}_1,\dots,\vec{x}_b) := \bigwedge_{i=1}^b g(\vec{x}_i) \), where each $\vec{x}_i$ has length $k$. Then, our lemma says that $\enc(g^{\land b}) = \widetilde{O}(\sqrt{b2^{k}})$. 

\begin{figure}[h]
\centering
\begin{tikzpicture}[
    x=1cm, y=1cm,
    font=\small,
    line cap=round,
    line join=round,
    >=Stealth
]
\def\chunkWidth{2.2}
\def\arrayHeight{0.65}
\pgfmathsetmacro{\cellWidth}{\chunkWidth/8}
\colorlet{querycolor}{red!70!black}

\foreach \left/\label/\cell in {0/1/1,2.2/2/5,5.1/i/3,8.0/b/6} {
    \pgfmathsetmacro{\right}{\left+\chunkWidth}
    \pgfmathsetmacro{\queryX}{\left+(\cell+0.5)*\cellWidth}

    \fill[black!3] (\left,0) rectangle (\right,\arrayHeight);
    \fill[querycolor!15]
        ({\left+\cell*\cellWidth},0) rectangle
        ({\left+(\cell+1)*\cellWidth},\arrayHeight);
    \foreach \j in {1,...,7} {
        \draw[black!25,thin]
            ({\left+\j*\cellWidth},0) --
            ({\left+\j*\cellWidth},\arrayHeight);
    }
    \draw[semithick] (\left,0) rectangle (\right,\arrayHeight);

    \draw[->,querycolor,semithick]
        (\queryX,1.12) node[above=2pt] {$\vec{x}_{\label}$}
        -- (\queryX,{\arrayHeight+0.04});

    \node[below=5pt] at ({(\left+\right)/2},0)
        {chunk $\label$};
    \draw[decorate,decoration={brace,mirror,amplitude=4pt}]
        (\left,-0.65) -- (\right,-0.65)
        node[midway,below=7pt] {$2^k/b$};
}

\foreach \left/\right in {4.4/5.1,7.3/8.0} {
    \draw[semithick] (\left,0) -- (\right,0);
    \draw[semithick]
        (\left,\arrayHeight) -- (\right,\arrayHeight);
    \node at ({(\left+\right)/2},{\arrayHeight/2}) {$\cdots$};
}

\node[left=6pt] at (0,{\arrayHeight/2}) {$A$};
\draw[decorate,decoration={brace,amplitude=5pt}]
    (0,1.75) -- (10.2,1.75)
    node[midway,above=8pt] {$2^k$ entries};
\end{tikzpicture}
\caption{The truth table of $g$, partitioned into $b$ chunks of $2^k/b$ entries each.}
\label{fig:array-chunks}
\end{figure}
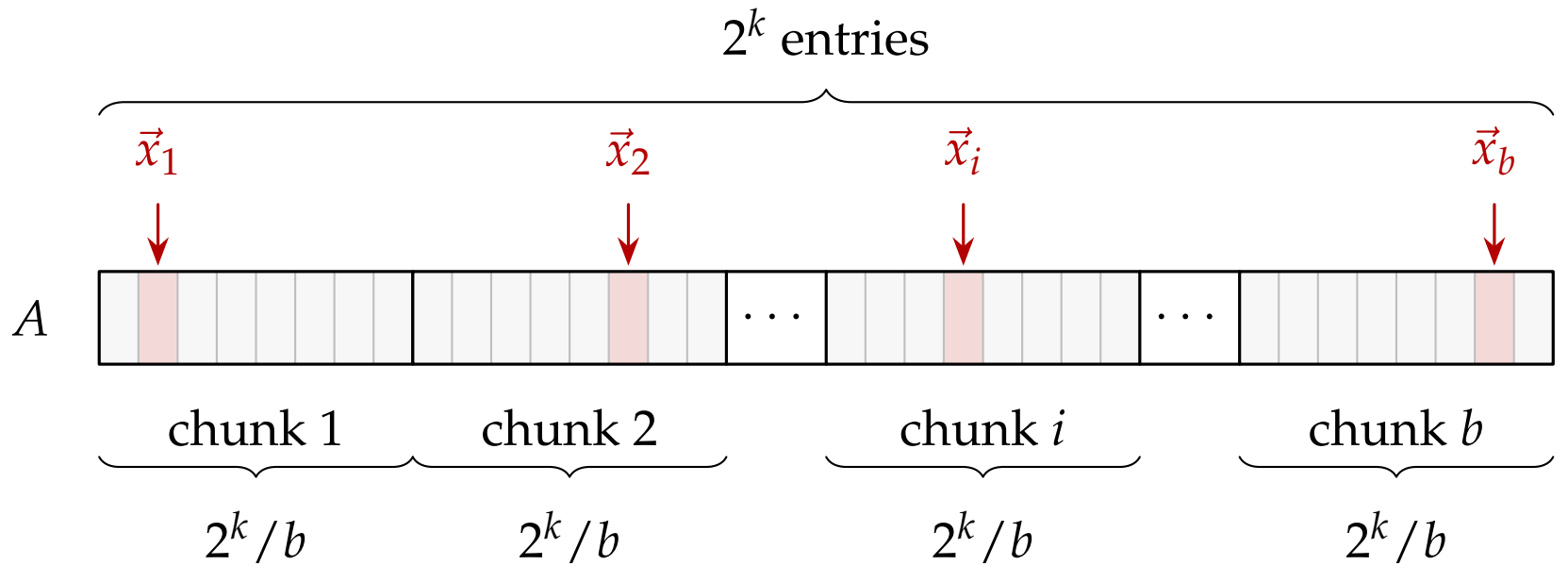

We think of the truth table of $g$ as an array $A$ of size $2^k$, indexed in binary by $\vec{x} \in \{\zero, \one\}^k$. Then, the key idea to prove~\Cref{thm:checkadvice} is that if we were promised that the inputs $\vec{x}_1, \ldots, \vec{x}_b$ on which we want to query the array $A$ were such that each $\vec{x}_i$ is in the $i$th $\frac{2^k}{b}$-sized chunk of $A$ (see~\Cref{fig:array-chunks}), then each $\vec{x}_i$ could be described with only $\lg (2^{k}/b)$ bits representing its position within the corresponding chunk, and thus the boolean function defined by whether those $k-\lg(b)$ bits indicate a $\one$ in $A$ can be encoded with $O(\sqrt{2^{k}/b})$ clauses according to~\Cref{thm:sparse}; taking the conjunction of these $b$ encodings would give $\enc(g^{\land b}) = b \cdot O(\sqrt{2^{k}/b}) = O(\sqrt{b2^k})$.

Naturally, we do not have this promise in general, but we can roughly simulate it by ``hashing'' each input $\vec{x}_i$ into $O(b)$ slots in such a way that the average number of possible inputs that are hashed into a slot is $\widetilde{O}(2^k/b)$. This hashing takes place within the encoding itself, using an explicit expander from Teutsch~\cite{Teutsch2014ShortLists}. As opposed to the simplified promise described above, where each input's address is indexed only by the $\lg(2^k / b)$ variables associated to its chunk, the inputs that are mapped into a given slot may have addresses anywhere in $A$. Nevertheless, we can still apply~\Cref{thm:sparse}, because each slot's function is \emph{sparse}, accepting only $\widetilde{O}(2^k/b)$ addresses on average (recall that for a function with $M$ satisfying inputs, the bound of~\Cref{thm:sparse} scales roughly with $\sqrt{M}$ rather than $\sqrt{2^n}$).\footnote{To keep the proof of~\Cref{thm:checkadvice} self-contained, we actually use the slightly weaker~\Cref{thm:sparse-simple}, which suffices here; the full proof of~\Cref{thm:sparse} is deferred to~\Cref{sec:sparse-ub}.}

We find it notable that sparsity enters the argument at all. The function $g$ we are querying is arbitrary and potentially dense, and both the na\"ive approach and the promise version above use only the special case $\enc(f) = O(\sqrt{2^n})$ for arbitrary $f$. The hashing step is what \emph{creates} sparsity: each of the $b$ queries may land anywhere in the dense table of $2^k$ entries, but hashing routes it to one of $O(b)$ slots, each responsible for only an $\widetilde{O}(1/b)$ fraction of the addresses on average. The dependence of~\Cref{thm:sparse} on the support size, in the regime $M = o(2^n)$, is exactly what lets us exploit this.

\subsection{\texorpdfstring{$k$}{k}-CNF boolean functions} \label{sec-overview-kcnf}

We start with a reduction, presented in~\Cref{sec-reduction-to-antitone}, that shows how upper bounds for encoding the independent-set function $\mathsf{IS}_G$ translate into upper bounds for general $k$-CNF functions. More precisely, if every $n$-vertex graph $G$ satisfies $\enc(\mathsf{IS}_G) \leq f(n)$ for some function $f$, then any $k$-CNF function can be encoded with $O_k(n^{k-2}f(n))$ clauses.

Thus, to prove~\Cref{thm-quasi}, it suffices to prove that $\enc(\mathsf{IS}_G) =  \frac{n^2}{2^{\Omega(\sqrt{\lg n})}}$ for all $n$-vertex graphs $G$. We do this by applying our~\Cref{thm:robson-advice} to a derandomized version of the algorithm of Larsen and Williams~\cite{larsenWilliams} for answering independent set queries. In order to derandomize their algorithm, we use a derandomization for the orthogonal vectors problem due to Chan and Williams~\cite{derandomized}. 

The proof of~\Cref{thm-polytime} explicitly constructs an encoding for $\mathsf{IS}_G$ without using~\Cref{thm:robson-advice}, which makes it much more technically sophisticated.  The reason for this is that we do not know how to save more than a $\lg^2 n$ factor within deterministic polynomial time, and since~\Cref{thm:robson-advice} loses polylogarithmic factors, its direct application could not beat the $n^k/\lg n$ simple known upper bound.
This proof, instead, is based on an efficient regularity-like graph decomposition~(\Cref{lem-decomp}), whose building blocks can be encoded compactly: 
bipartite graphs without large empty squares (i.e., their bipartite complement does not have large bicliques), and a single sparse subgraph. The sparse part can be encoded directly by known results on biclique partitions, while the bipartite graphs without large empty squares are handled by a dedicated encoding which uses hashing (via explicit expanders) in a similar way to~\Cref{thm:robson-advice}. Computing this decomposition in deterministic polynomial time is technically challenging, and relies on an efficient algorithm for identifying sparse squares in a graph (\Cref{lem-sparse-rectangle-search}) which is based on derandomizing a random walk via explicit expanders in the same spirit as the derandomization of graph products from Alon, Feige, Wigderson, and Zuckerman~\cite{DBLP:journals/cc/AlonFWZ95}.

Finally, in \Cref{sec-kcnf-lb}, we prove our lower bounds for $k$-CNF functions using information-theoretic arguments, which are the only technique we have for proving superlinear lower bounds on encodings. For encodings of bounded width, a direct counting argument gives the $\Omega_k(n^k/\lg n)$ lower bound mentioned above. Without a width bound, we need to bound the number of minimal encodings with at most $q$ clauses, which in turn requires a bound on the number of auxiliary variables that can appear in such an encoding. We prove in \Cref{prop:auxiliaries} that a minimal encoding has fewer auxiliary variables than clauses. Intuitively, if we have more auxiliary variables than clauses in an encoding, then clauses become ``too easy'' to satisfy. To formalize this idea, we show using Hall's marriage theorem that if there are at least as many auxiliary variables as clauses, then there is a partial assignment $\alpha$ to the auxiliary variables that satisfies every clause containing a variable in the domain of $\alpha$. Such an $\alpha$ is called an \emph{autarky}~\cite{monienSolvingSatisfiabilityLess1985}, and it is a straightforward argument to show that an encoding with an autarky cannot be minimal. This yields \Cref{lem-counting}, which states that at most $3^{(n+q)q}+1$ functions on $n$ inputs have encodings with at most $q$ clauses; this lemma gives the $\Omega_k(n^{k/2})$ bound of \Cref{prop-kcnf-unbounded-width}, as well as the lower bounds for sparsely supported functions described next.

\subsection{Sparsely supported functions}\label{subsec:sparse}

In \Cref{sec:arbitrary}, we prove \Cref{thm:sparse}, which states that the worst-case encoding complexity of a function $f \colon \{\zero,\one\}^n \to \{\zero,\one\}$ with exactly $M$ satisfying inputs is $\enc(n,M) = \Theta(n+\sqrt{M(n-\lg M)})$ for $1 \le M \le 2^{n-1}$.
Our upper-bound construction translates a succinct static dictionary~\cite{BrodnikMunro1999,Pagh2001,Patrascu2008,HuLYZZ2025} directly into an encoding. This further illustrates the principle behind \Cref{thm:robson-advice} of translating data structures into encodings; perhaps remarkably, here an asymptotically space-optimal data structure yields an asymptotically optimal encoding.

Conceptually, the dictionary reduces the sparse case to the dense case: it compresses $f^{-1}(\one)$ into a table $\sigma_f$ of $S = O(M(n-\lg M))$ bits, from which $f$ is evaluated with $O(1)$ lookups and word operations. Here, each of the $O(1)$ word operations can be encoded directly via the Tseitin transformation (without even invoking Robson's theorem~\cite{robson1991ot}), so the main difficulty lies in the lookups. It is tempting to handle each lookup by viewing $\sigma_f$ as an arbitrary function on $\lg S$ variables and applying a construction for arbitrary functions with $O(\sqrt{S})$ clauses, such as~\Cref{prop:dense} or Nechiporuk's nondeterministic branching programs~\cite{nechiporuk1962zero}. However, each lookup retrieves an entire word of $w = \Theta(n)$ bits, and fetching it one bit at a time costs $O(w\sqrt{S})$ clauses, a factor of $\Theta(n)$ too many.

We overcome this with~\Cref{lem:sparse-word-table}, which may be of independent interest: an entire block of $w$ bits of a table of size $S$ can be fetched with $O(\sqrt{S}+w)$ clauses. Since even a single bit of a worst-case table requires $\Omega(\sqrt{S})$ clauses (\Cref{prop:dense}), the remaining bits come essentially for free when $w \le \sqrt{S}$. Checking each memory read with this lemma then gives the upper bound of \Cref{thm:sparse}.

In turn, we derive \Cref{lem:sparse-word-table} from \Cref{thm:sparse-simple}, a slightly weaker version of~\Cref{thm:sparse} with $O(n+\sqrt{nM})$ clauses that we prove directly in~\Cref{sec:sparse}. Its proof arranges the $M$ satisfying inputs in an $R \times C$ table and guesses an entry, with the column written in binary and the row chosen by one-hot variables. Rather than checking each row separately, which would cost $n$ clauses per row, for each column, each value of each input bit $x_i$ implies the disjunction of the rows whose entry agrees with it. Thus we pay only for the $C$ columns and $n$ coordinates, giving $O(R+nC) = O(n+\sqrt{nM})$ clauses for $C = \Theta(\sqrt{M/n})$.
\Cref{lem:sparse-word-table} is then a direct application of~\Cref{thm:sparse-simple} to the function that accepts an address $j$ and a word $\vec{z}$ if and only if $\vec{z}$ is stored at address $j$; this function has $\Theta(w)$ variables and $S/w$ satisfying inputs, giving $O(w+\sqrt{w\cdot S/w}) = O(\sqrt{S}+w)$ clauses. In other words, the dictionary turns the sparse problem into a dense one, and~\Cref{lem:sparse-word-table} turns it back into a sparse one, loosely echoing how hashing creates sparsity in the proof of~\Cref{thm:checkadvice}.

\section{Encoding Nonuniform Computation}\label{sec:non-uniform}
In this section, we prove \Cref{thm:robson-advice}:
\robsonadvice*

Our proof relies on several ingredients, including a modification to Robson's theorem~\cite{robson1991ot}.
Robson proved that given a nondeterministic RAM machine $\mathcal{M}$ and an input $\vec{x}$ such that $\mathcal{M}(\vec{x})$ runs in time $t$, one can construct in deterministic polynomial time a CNF formula $\varphi_{\mathcal{M}, \vec{x}}$ of size $O(t \lg t)$ such that $\mathsf{SAT}(\varphi_{\mathcal{M}, \vec{x}})$ if and only if $\mathcal{M}(\vec{x})$ accepts. However, by inspecting the proof it is easy to see that his construction is ``uniform'' over the inputs, meaning that a very slight modification yields a formula $\varphi_{\mathcal{M}}$, with $\vec{x}$ as the input variables (and with auxiliary variables), and such that $\mathcal{M}(\vec{x})$ accepts if and only if $\mathsf{SAT}(\varphi_{\mathcal{M}}(\vec{x}))$. Thus, we have the following:

\begin{theorem}[Robson's theorem~\cite{robson1991ot}]\label{thm:robson-f}
    Let $f \colon \{\zero,\one\}^n \to \{\zero,\one\}$, and suppose that $f$ can be evaluated in time $t$ by a nondeterministic {\emph{RAM}} machine. Then, $\enc(f) = O(t \lg t)$. Furthermore, such an encoding can be constructed deterministically in time polynomial in $n+t$.
\end{theorem}

\subsection{Nondeterministic RAM with advice (ANRAM)}
Our nondeterministic-RAM-with-advice model closely follows that of the nondeterministic RAM used by \cite{robson1991ot}. In Robson's model, a machine is specified by a sequence of instructions $(i_1, \ldots, i_T)$, which are to be executed sequentially by default, but include the possibility of jumping to other instructions, as for example $i_7 := \text{JUMP } 3$, which would make the machine go back to execute $i_3$. For the sake of our asymptotic results, we follow Robson in assuming that the number of instructions is upper-bounded by a fixed constant; this assumption is without loss of generality, because we can use a universal machine and encode the instructions as part of the advice. The input is given on a read-only tape, which is read sequentially (i.e., no random access). 
There are infinitely many registers $r_0, r_1, r_2,\ldots$, with $r_0$ representing a special accumulator register where operations take place. The machine has instructions for loading data into the accumulator, storing the contents of the accumulator, adding and subtracting to the accumulator, unconditional jumping, conditional jumping, nondeterministic jumping, and halting. As with Robson, we use the \emph{logarithmic cost criterion}, where the cost of an operation is proportional to the lengths of its operands measured in bits.

The above model of computation constitutes nondeterministic RAM. For ANRAM, we allow the machine to make random-access queries to a read-only advice memory containing an advice string $A$, which does not depend on the input.
In principle, such a machine could query the advice string at any point throughout a computation. It will be convenient, however, to assume all advice queries are made at the very beginning. Thanks to nondeterminism, this assumption can be made without loss of generality, since such a machine can nondeterministically guess, at the beginning of a computation, which advice queries will be made during the remaining computation. After making these queries, the machine operates as a nondeterministic RAM machine \emph{without advice}, which uses the prefetched advice queries instead of directly querying the advice memory. If the machine ever needs the $i$th bit of advice, and it is not available in the registers corresponding to prefetched advice queries, then it simply rejects. Formally, we stipulate that an ANRAM machine operates as follows after receiving an input $\vec{x}$:
\begin{enumerate}
    \item First, the machine nondeterministically guesses the indices $i_1, i_2, \dots, i_b$ of $A$ that will be queried during the computation on $\vec{x}$. The machine then fetches those bits of advice, storing the results as a list of ordered pairs $((i_1, A[i_1]), \dots, (i_b, A[i_b]))$.
    \item Then, the machine passes $\vec{x}$ together with the advice queries $((i_1, A[i_1]), \dots, (i_b, A[i_b]))$ to a nondeterministic RAM machine without advice.
\end{enumerate}

\subsection{Proof of~\texorpdfstring{\Cref{thm:robson-advice}}{the advice-to-CNF theorem}}

To prove~\Cref{thm:robson-advice}, we will use auxiliary variables that serve to nondeterministically guess the advice string, and then
use~\Cref{thm:robson-f} in a black-box fashion. The main technical difficulty, however, is that we cannot afford to guess the entire advice string, but only the $b$ bits that will be accessed during the computation, and thus we will need to encode into CNF a ``check-advice'' subroutine that ensures that the $b$ bits guessed, which can vary depending on the input, are compatible with the global advice string $A$. 
It turns out that the subroutine we need for checking advice is a particular case of a more general problem: obtaining compact encodings for ``multiple queries'' to a boolean function. The following lemma, in the spirit of ``mass production theorems''~\cite{Ulig1974,Kretschmer23}, shows that querying a boolean function $b$ times on arbitrary inputs can be encoded with substantially fewer clauses than $b$ independent encodings.

\begin{lemma} \label{thm:checkadvice}
    Let $f\colon \{\zero,\one\}^n \rightarrow \{\zero,\one\}$, and let $b \in [2^n]$. Let $f^{\land b} \colon \{\zero,\one\}^{bn} \rightarrow \{\zero,\one\}$ be such that for every $\vec{x_1}, \ldots, \vec{x_b} \in \{\zero, \one\}^n$ we have 
    \[ 
        f^{\land b}(\vec{x_1}, \ldots, \vec{x_b}) = \bigwedge_{i \in [b]} f(\vec{x_i}).
    \]
    Then, $\enc(f^{\land b}) = \widetilde{O}(\sqrt{2^n b})$. Furthermore, such an 
    encoding can be obtained deterministically in time polynomial in $2^n$ and $b$.
\end{lemma}

\noindent
Before proving \Cref{thm:checkadvice}, we show how it yields \Cref{thm:robson-advice}.

\begin{proof}[Proof of \Cref{thm:robson-advice}]
    Let $f \colon \{\zero,\one\}^n \to \{\zero,\one\}$ be a boolean function, and suppose that $\mathcal{A}$ is an ANRAM machine that evaluates $f$ in time $t$ with an advice string $A$ consisting of $m$ bits such that at most $b$ bits of $A$ are queried for each input. The cases $m=0$ or $b=0$ follow directly from \Cref{thm:robson-f}, so assume $1\leq b\leq m$. Pad $m$ to a power of two, increasing it by a factor of at most two. Also, if fewer than $b$ bits are read, it is always possible to read additional bits without changing the computation. Thus, we assume exactly $b$ bits of advice are read per input.
    
  We construct our encoding of $f$ as follows. Introduce $\vec{i_1}, \dots, \vec{i_b}$, where each $\vec{i_j}$ is a sequence of $\lg m$ auxiliary variables that represent addresses of $A$. Furthermore, we have auxiliary variables
    $a_1, \dots, a_b$, representing the values of the fetched bits of advice; intuitively, we want to enforce that $A[\vec{i_j}] = a_j$.
    Let $h\colon \{\zero, \one\}^{\lg m + 1} \to \{\zero, \one\}$ be the boolean function that takes the binary expansion of a number $j \in [0, m-1]$ and a value $v$, and returns whether $A[j] = v$. Then, by invoking~\Cref{thm:checkadvice} with $h$, there is an encoding 
    \[
        (\psi, (\vec{i_1}, a_1, \dots, \vec{i_b}, a_b), Y)
    \]
    with $\widetilde{O}(\sqrt{mb})$ clauses such that $\psi$ is satisfiable for an assignment $\tau$ to its input variables if and only if
    \[
    \forall j \in [b], \quad A\big[\tau(\vec{i_j})\big] = \tau(a_j).
    \]

    According to our definition of an ANRAM machine, $\mathcal{A}$ invokes a nondeterministic RAM machine (without advice) $\mathcal{B}$ on an input of length $n' := n + b \cdot (\lg(m)+1)$. Let $g \colon \{\zero,\one\}^{n'} \to \{\zero,\one\}$ be the boolean function computed by $\mathcal{B}$. 
    By \Cref{thm:robson-f}, there is an encoding $\varphi_{\mathcal{B}}$ of $g$ with the $n'$ input variables $\vec{x}, \vec{i_1}, a_1, \dots, \vec{i_b}, a_b$ and using $\widetilde{O}(t+n') = \widetilde{O}(t+b \lg m)$ clauses. 
     Then, our encoding for $f$ is simply $\psi \land \varphi_{\mathcal{B}}$, which has $\widetilde{O}(t + \sqrt{mb})$ clauses. The correctness follows from the correctness of its components. Indeed, if $f(\vec{x}) = \one$, then there is some assignment $\tau$ to $\vec{i_1}, \dots, \vec{i_b}, a_1, \dots, a_b$ representing the advice queries performed by $\mathcal{A}$ and fed to $\mathcal{B}$, and then $\varphi_\mathcal{B}$ is satisfiable for this assignment since $\mathcal{B}$ accepts on input $(\vec{x}, \tau(\vec{i_1}), \tau(a_1), \dots, \tau(\vec{i_b}), \tau(a_b))$. Conversely, if $\tau$ is a satisfying assignment to $\psi \land \varphi_\mathcal{B}$, then by \Cref{thm:robson-f}, $\mathcal{B}$ accepts on $(\vec{x}, \tau(\vec{i_1}), \tau(a_1), \dots, \tau(\vec{i_b}), \tau(a_b))$.
   By~\Cref{thm:checkadvice}, $A\big[\tau(\vec{i_j})\big] = \tau(a_j)$, so $\mathcal{B}$ accepts on 
    $(\vec{x}, \tau(\vec{i_1}), A[\tau(\vec{i_1})], \dots, \tau(\vec{i_b}), A[\tau(\vec{i_b})])$. Therefore, $\mathcal{A}$ accepts on $\vec{x}$ and we conclude
    $f(\vec{x}) = \one$.

    Finally, note that the deterministic polynomial-time construction follows directly from the deterministic polynomial-time constructions of \Cref{thm:checkadvice} and \Cref{thm:robson-f}.
\end{proof}

\subsection{Encoding sparse functions}\label{sec:sparse}
To check the advice queries, we need compact encodings of functions with few satisfying inputs. Our construction uses a compact encoding for the function $\amo(x_1, \ldots, x_n)$, which evaluates to $\one$ if and only if at most one of the variables $x_1, \ldots, x_n$ is $\one$. The following is the best known upper bound on $\enc(\amo(x_1, \ldots, x_n))$:
\begin{proposition}[\cite{NearOptimalCardinalitySubmitted}]\label{prop:amo-bound}
    We have $\enc(\amo(x_1, \ldots, x_n)) \leq 2n + 2\sqrt{2n} + \widetilde{O}(\sqrt[4]{n})$.
\end{proposition}

\noindent
The following lemma bounds the encoding complexity of sparsely supported functions.

\begin{lemma}\label{thm:sparse-simple}
    Let $f\colon \{\zero,\one\}^n\to\{\zero,\one\}$, and let $M:=|f^{-1}(\one)|\ge1$. Then,
    \[
        \enc(f)=O\left(n+\sqrt{nM}\right).
    \]
    Given the list of elements of $f^{-1}(\one)$, an encoding attaining this bound can be constructed deterministically in time polynomial in $nM$.
\end{lemma}
\begin{proof}
    Arrange the elements of $f^{-1}(\one)$ in an $R\times C$ table $T$, where $C$ is a power of two to be specified later and $R:=\ceil{M/C}$. Pad the table by repeating any element of $f^{-1}(\one)$. For $r\in[0,R-1]$ and $v\in[0,C-1]$, write $T[r,v]\in\{\zero,\one\}^n$ for the entry in row $r$ and column $v$, and $T[r,v]_i$ for its $i$th coordinate. Introduce a vector $\vec{c}$ of $\lg C$ auxiliary variables to select a column, and auxiliary variables $y_0,\ldots,y_{R-1}$ to select a row. Enforce exactly one true row variable using an encoding of $\amo(y_0,\ldots,y_{R-1})$ and the clause $\bigvee_{r=0}^{R-1} y_r$. This requires $O(R)$ clauses by \Cref{prop:amo-bound}. We note that the encoding from \Cref{prop:amo-bound} can be constructed in deterministic polynomial time.

    For each column $v\in[0,C-1]$, let $D_v(\vec{c})$ be the clause asserting that $\vec{c}$ is not the binary representation of $v$. For every $v$ and every coordinate $i\in[n]$, add the two clauses
    \begin{align*}
        D_v(\vec{c})\lor\overline{x_i}\lor
            \bigvee_{\substack{r \\ T[r,v]_i=\one}}y_r \\
        D_v(\vec{c})\lor x_i\lor
            \bigvee_{\substack{r \\ T[r,v]_i=\zero}}y_r.
    \end{align*}

    Fix a satisfying assignment, write $\vec{x}$ for its values on the input variables, let $v\in[0,C-1]$ be the number represented by $\vec{c}$, and let $r$ be the unique index with $y_r=\one$. Then $D_v(\vec{c})$ is false, so for every $i$, the two clauses for $v$ and $i$ force $x_i=T[r,v]_i$; that is, $\vec{x}=T[r,v]\in f^{-1}(\one)$. Conversely, suppose $\vec{x}\in f^{-1}(\one)$. Since every element of $f^{-1}(\one)$ appears in the table, we have $\vec{x}=T[r,v]$ for some $r\in[0,R-1]$ and $v\in[0,C-1]$. Set $\vec{c}$ to the binary representation of $v$, set $y_r=\one$ and every other row variable to $\zero$, and extend this assignment to the auxiliary variables of the encoding of $\amo(y_0,\ldots,y_{R-1})$. If $T[r,v]_i=\one$, the first clause for $v$ and $i$ contains $y_r$, which is true, and the second contains $x_i$, which is true since $x_i=T[r,v]_i$; the case $T[r,v]_i=\zero$ is symmetric. The clauses for every other column $v'$ are satisfied by $D_{v'}(\vec{c})$. Thus, the encoding is satisfiable when $\vec{x}\in f^{-1}(\one)$. This completes the proof that the encoding is correct.

    The number of clauses is $O(R+nC)$. If $M\ge n$, choose a power of two $C=\Theta(\sqrt{M/n})$; if $M<n$, choose $C=1$. In both cases the bound is $O(n+\sqrt{nM})$. All table entries and clauses can be written in time polynomial in $nM$.
\end{proof}

In \Cref{thm:sparse}, proven in \Cref{sec:arbitrary}, we sharpen this result to $O(n+\sqrt{M(n-\lg M)})$ clauses for $1\le M\le2^{n-1}$, and prove that this is optimal in the worst case up to constant factors.

\subsection{Proof of \texorpdfstring{\Cref{thm:checkadvice}}{the multi-query lemma}}

For the proof of \Cref{thm:checkadvice}, we need to encode a kind of hash table, so in order to keep the encoding constructible in \emph{deterministic} polynomial time, we will make use of explicit expanders. Intuitively, we represent a hash table by a bipartite incidence graph: left vertices are \emph{universe elements} and right vertices are \emph{slots}, and a form of Hall's condition~\cite{hall} satisfied by the considered expander will guarantee that any sufficiently small set of universe elements can be matched into distinct slots. Concretely, we will use the following lemma.

\begin{lemma}[{Teutsch's expander lemma~\cite[Section 4]{Teutsch2014ShortLists}}]
\label{lem:teutsch}
For every integer $k\geq0$ and finite universe $L$ with $|L|\geq2^k$, there is a bipartite graph $(L,R,E)$ with $|R|<2^{k+3}$, left degree bounded by a fixed polynomial in $\lg |L|$, and $|N(S)|\geq|S|$ for every $S\subseteq L$ of size at most $2^k$. For each left vertex, its $i$th neighbor can be computed deterministically in time polynomial in $\lg|L|$ and $\lg d$, where $d$ is the maximum left degree.
\end{lemma}

\begin{proof}[Proof of \Cref{thm:checkadvice}]
     Let $k:=\ceil{\lg b}$ and $q := 2^{k+3}$. Apply \Cref{lem:teutsch} to $L:=\{\zero,\one\}^n$ with this $k$, and pad the right vertex set with isolated vertices until it is $[q]$. Write the resulting graph as $G=(L,[q],E)$.
    For $r \in [q]$, define $g_r\colon\{\zero,\one\}^n\to\{\zero,\one\}$ by
    \[
        g_r(\vec{y}):=f(\vec{y})\land[(\vec{y},r)\in E].
    \]
    Since $|g_r^{-1}(\one)|\leq \deg(r)$, \Cref{thm:sparse-simple} gives a formula $\varphi_r$ encoding $g_r$ with $O(n+\sqrt{n\deg(r)})$ clauses. If $g_r^{-1}(\one)=\emptyset$, take $\varphi_r$ to consist of the empty clause. 

    Introduce vectors of auxiliary variables $\vec{y}_1,\ldots,\vec{y}_q$ each of length $n$, and individual auxiliary variables $a_1,\ldots,a_q$, representing the contents of the slots and whether they are occupied or not, respectively. Let $h$ be the boolean function defined by
    \[
    h(\vec{x}_1,\ldots,\vec{x}_b,\vec{y}_1,a_1,\ldots,\vec{y}_q,a_q)
        :=\bigwedge_{i\in[b]}\bigvee_{r\in[q]}
        \bigl(a_r\land[\vec{x}_i=\vec{y}_r]\bigr).
    \]
    The fact that $\enc(h) = \widetilde{O}(nb)$ follows directly from Robson's result since an NRAM can evaluate $h$ in time $\widetilde{O}(nb)$: for each $i \in [b]$ guess an index $r_i \in [q]$, and check that both $a_{r_i} = \one$ and $\vec{x}_i = \vec{y}_{r_i}$. 
    Thus, by \Cref{thm:robson-f}, $h$ has an encoding $\psi$ with $\widetilde{O}(nb)$ clauses. 
    
    Our encoding of $f^{\land b}$ is now given by
    \[
        \Phi:=\psi\land\bigwedge_{r\in[q]}\ \bigwedge_{C\in\varphi_r}
        (\overline{a_r}\lor C),
    \]
    and we assume that the sets of auxiliary variables internal to $\psi$ and each of the $\varphi_r$ are pairwise disjoint.

    To prove correctness, suppose first that an assignment $\tau$ satisfies $f(\tau(\vec{x}_i))=\one$ for every $i\in[b]$. Then, by~Hall's theorem~\cite{hall}, the set $S := \{ \tau(\vec{x}_i) \mid i\in[b] \}$, which has size at most $b \leq 2^k$, admits a matching $M$ in $G$ saturating $S$ (i.e., an injective mapping $M \colon S \to [q]$ with $(s, M(s)) \in E$ for every $s \in S$). 
    Extend $\tau$ by an assignment $\theta$ such that  $\theta(a_r)=\one$
    if and only if $r$ is in the image of $M$, and $\theta(\vec{y}_r) :=\tau(\vec{x}_i)$ for every $i\in[b]$ such that $M(\tau(\vec{x}_i))=r$. Then $h$ evaluates to $\one$ under $\theta$, and every occupied slot $r$ (i.e., $\theta(a_r)=\one$) satisfies $g_r(\theta(\vec{y}_r))=\one$. Thus, the auxiliary variables of $\psi$ and the $\varphi_r$ can be chosen to satisfy $\Phi$. Conversely, any satisfying assignment $\theta$ to $\Phi$ gives, for each $i\in[b]$, a slot $r$ with $\theta(a_r)=\one$ and $\theta(\vec{x}_i)=\theta(\vec{y}_r)$. The clauses $(\overline{a_r} \lor C)$ for $C\in\varphi_r$ force $\theta$ to satisfy $\varphi_r$, which in turn implies $g_r(\theta(\vec{y}_r))=\one$, and hence $f(\theta(\vec{x}_i))=\one$. 
    
    Let us now bound the number of clauses. First,
    \begin{align*}
        |\Phi| = |\psi| + \sum_{r \in [q]}  |\varphi_r| = \widetilde{O}(nb) + \sum_r O\left(n+\sqrt{n\deg(r)}\right).    
    \end{align*}
    Then, by the Cauchy--Schwarz inequality, and letting $d := \max_{\ell \in L} \deg(\ell)$,
    \[
        \sum_{r\in[q]}\sqrt{n \deg(r)}
        \leq\sqrt{nq\sum_{r\in[q]} \deg(r)}
        \leq\sqrt{nq\,2^n d}.
    \]
    Therefore, the total number of clauses is
    \[
        \widetilde{O}(nb)+O\bigl(nq+\sqrt{nq\,2^n d}\bigr)
        =\widetilde{O}(\sqrt{2^n b}),
    \]
    since $q=O(b)$, $b\leq2^n$, and $d=\operatorname{poly}(n)$.

    Finally, we can explicitly construct $G$ in time polynomial in $2^n$ by~\Cref{lem:teutsch}, and then enumerate the satisfying assignments of every $g_r$ within the same time bound. Thus, each $\varphi_r$ can be computed in deterministic time $\operatorname{poly}(2^n)$ by~\Cref{thm:sparse-simple}, and $\psi$ can be computed deterministically in time $\operatorname{poly}(2^n)$ by~\Cref{thm:robson-f}. Thus, $\Phi$ can be computed deterministically in time $\operatorname{poly}(2^n)$.
\end{proof}

\section{Encoding \texorpdfstring{$k$}{k}-CNF Functions}\label{sec:kcnf}
In this section, we prove \Cref{thm-quasi,thm-polytime}. Both theorems reduce to compactly encoding $\mathsf{IS}_G$, where $G$ is a bipartite graph. We prove \Cref{thm-quasi} by showing that there is a deterministic data structure for computing $\mathsf{IS}_G$ that uses $O(n^2)$ bits of memory whose query time is $n^2/2^{\Omega(\sqrt{\lg n})}$. This is a derandomization of a data structure of Larsen and Williams~\cite{larsenWilliams}, which we give in \Cref{sec:derandomization}. Together with \Cref{thm:robson-advice}, this proves \Cref{thm-quasi}. For \Cref{thm-polytime}, we encode sparse graphs using biclique partitions in \Cref{sec-sparse-graphs}, and bipartite graphs with small bipartite independence number using hash tables in \Cref{sec-small-abip}. In \Cref{sec-constructive-graphs}, we decompose an arbitrary bipartite graph into pieces to which one of these two encodings applies. We conclude with counting lower bounds in \Cref{sec-kcnf-lb}.

\subsection{Derandomizing the Larsen--Williams data structure}\label{sec:derandomization}
We adapt the strategy of Larsen and
Williams~\cite{larsenWilliams}, replacing its randomized orthogonal-vector
subroutine from Abboud, Williams, and Yu~\cite{DBLP:conf/soda/AbboudWY15} by the deterministic algorithm of Chan and
Williams~\cite{derandomized}. Following Larsen and Williams, we formulate our goal as preprocessing a boolean matrix $A$
to compute queries of the form $u^TAv$ over the boolean semiring, since this is equivalent to querying $\mathsf{IS}_G$ for a bipartite graph $G$.

\begin{theorem}\label{thm:derandomized}
For every boolean matrix $A\in\{\zero,\one\}^{n\times n}$, there is a data
structure using $O(n^2)$ bits of memory for which, given
$u,v\in\{\zero,\one\}^n$, one can deterministically compute
$u^TAv$ over the boolean semiring in time
$n^2/2^{\Omega(\sqrt{\lg n})}$.
\end{theorem}

For $U,V\subseteq[n]$, we use notation
\(
 A\{U\times V\}:=\{(i,j)\in U\times V:A(i,j)=\one\}.
\)
If $U$ and $V$ are the supports of $u$ and $v$, then $u^TAv=\one$
if and only if $A\{U\times V\}\neq\varnothing$.

\subsubsection{Preprocessing}

Call a rectangle $U\times V$ \emph{empty} if $A\{U\times V\}=\varnothing$.
Given a list $L$ of rectangles, let
\[
 C(L):=[n]^2\setminus\bigcup_{(U,V)\in L}(U\times V).
\]

\begin{lemma}\label{claim:L}
For every $R\geq1$, there is a list $L$ of at most $R$ empty
rectangles such that every empty rectangle $U\times V$ satisfies
\[
 |(U\times V)\cap C(L)|<n^2/R.
\]
\end{lemma}
\begin{proof}
Start with $L=\varnothing$.  While some empty rectangle has at least
$n^2/R$ entries in $C(L)$, add it to $L$.  Each addition removes at
least $n^2/R$ entries from $C(L)$, so there are at most $R$ additions.
\end{proof}

Choose constants $\varepsilon,\alpha>0$ with $3\alpha<\varepsilon$
as in \Cref{thm:black-boxed} below, and set
$R:=2^{\varepsilon\sqrt{\lg n}}$.
Construct $L$ as above, and enumerate it as
$(U_1,V_1),\ldots,(U_{|L|},V_{|L|})$.
Store $A$ as a bit matrix, and for every row $i$ and column $j$, store
vectors $a_i,b_j\in\{\zero,\one\}^{|L|}$ such that for every $r\in[|L|]$,
\[
 a_{i,r}:=
 \begin{cases}
  \one & \text{if }i\in U_r,\\
  \zero & \text{otherwise},
 \end{cases}
 \qquad
 b_{j,r}:=
 \begin{cases}
  \one & \text{if }j\in V_r,\\
  \zero & \text{otherwise}.
 \end{cases}
\]
Note that these vectors satisfy
\begin{equation}\label{eq:derand-orthogonality}
 (i,j)\in C(L)\quad\Longleftrightarrow\quad
 \langle a_i,b_j\rangle=0.
\end{equation}
Furthermore, the total space used is $O(n^2)$ bits.

\subsubsection{Answering queries}\label{subsec:answering}

For a query $(U,V)$, let $W:=(U\times V)\cap C(L)$. Since the rectangles of $L$ are empty, every $\one$-entry of $A$ lies
in $C(L)$. Thus, $A\{U\times V\}\subseteq W$, so our goal is to compute whether $W$ contains a $\one$. We proceed as follows.
\begin{enumerate}
\item Compute $|W|$ as described below.  If $|W|\geq n^2/R$, return
$\one$.
\item Otherwise, determine as described below whether
$A(i,j)=\one$ for some $(i,j)\in W$, and return the result.
\end{enumerate}
If $U\times V$ is empty, then $|W|<n^2/R$ by \Cref{claim:L}, so a
return in step 1 is correct.  Step 2 is correct because
$A\{U\times V\}\subseteq W$.  Moreover, whenever step 2 is executed, we have $|W|<n^2/R$. It remains to show that both steps of the algorithm can be computed in time $n^2/2^{\Omega(\sqrt{\lg n})}$.

\subsubsection{Counting the uncovered entries}\label{subsec:W}

By \eqref{eq:derand-orthogonality}, $W$ consists of the pairs
$(i,j)\in U\times V$ with $\langle a_i,b_j\rangle=0$.  Thus, both steps
of the query algorithm amount to an orthogonal-vectors problem on the
vectors $a_i$ for $i\in U$ and $b_j$ for $j\in V$, which we solve with
the deterministic algorithm of Chan and Williams~\cite{derandomized};
Larsen and Williams~\cite{larsenWilliams} used essentially the same
subroutine, with the randomized algorithm of Abboud, Williams, and
Yu~\cite{DBLP:conf/soda/AbboudWY15} instead.  Rather than only the total
number of orthogonal pairs, we need the number of orthogonal pairs
between every pair of blocks of partitions of $U$ and $V$.

\begin{lemma} \label{thm:black-boxed}
There are constants $\varepsilon,\alpha>0$ with $3\alpha<\varepsilon$
for which the following holds.  Let
$\beta:=\lfloor2^{\alpha\sqrt{\lg n}}\rfloor$.
Let $x_1,\ldots,x_n$ and $y_1,\ldots,y_n$ be boolean vectors of
dimension at most $2^{\varepsilon\sqrt{\lg n}}$, and let
$I_1,\ldots,I_p$ and $J_1,\ldots,J_q$ be subsets of $[n]$, each of
size at most $\beta$, with $p,q\leq\lceil n/\beta\rceil$.  Then all counts
\[
 Z_{h,\ell}:=|\{(i,j)\in I_h\times J_\ell:\langle x_i,y_j\rangle=0\}|
\]
can be computed deterministically in time
$\widetilde O(n^2/2^{\alpha\sqrt{\lg n}})$.
\end{lemma}
\begin{proof}
This is a minor modification of the algorithm of Chan and Williams~\cite[Theorem~1.3]{derandomized} for counting orthogonal pairs. Given $N$ vectors of dimension $c\lg N$, where $c\leq2^{\delta\sqrt{\lg N}}$ for a sufficiently small constant $\delta>0$, they deterministically count the orthogonal pairs of distinct vectors. To do so, they partition the vectors into blocks of at most $s:=N^{1/(k\lg c)}$ vectors each, for a sufficiently large constant $k$, and construct a polynomial whose value on the vectors of two blocks determines the number of orthogonal pairs between them. They evaluate this polynomial on all pairs of blocks at once with a single rectangular matrix multiplication, and add up the resulting counts. We change this in two ways. First, the vectors come from two collections: one factor of the matrix multiplication is built from the blocks $I_h$ of the vectors $x_i$, and the other from the blocks $J_\ell$ of the vectors $y_j$. Second, we output the count for every pair of blocks instead of their sum.
For sufficiently small $\varepsilon$ and $\alpha:=\varepsilon/4$, our vectors have dimension in their range and our blocks have at most $s$ elements, so the same analysis computes all counts $Z_{h,\ell}$ in time $\widetilde O(n^2/\beta^2)$.
\end{proof}

\begin{lemma}\label{lem:compute-W}
Both steps of the query algorithm can be carried out deterministically
in time $n^2/2^{\Omega(\sqrt{\lg n})}$.
\end{lemma}
\begin{proof}
Partition $U$ into $p:=\lceil|U|/\beta\rceil$ sets $I_1,\ldots,I_p$ and $V$
into $q:=\lceil|V|/\beta\rceil$ sets $J_1,\ldots,J_q$, each of size at most
$\beta=\lfloor2^{\alpha\sqrt{\lg n}}\rfloor$.
Apply \Cref{thm:black-boxed} with $x_i:=a_i$ and $y_j:=b_j$; this is
possible because these vectors have dimension
$|L|\leq R=2^{\varepsilon\sqrt{\lg n}}$.
By \eqref{eq:derand-orthogonality}, its count $Z_{h,\ell}$ is exactly
$|W\cap(I_h\times J_\ell)|$, so $|W|=\sum_{h,\ell}Z_{h,\ell}$.
This gives step 1 in time $\widetilde O(n^2/2^{\alpha\sqrt{\lg n}})$.

For step 2, check all entries of every block pair
$I_h\times J_\ell$ with positive count, and return $\one$ precisely
when some entry satisfies $A(i,j)=\one$. The block products are pairwise disjoint, so at most $|W|$
of them have positive count.  Since step 2 is executed only if $|W|<n^2/R$, the number
of entries checked is at most
\[
 |W|\beta^2<\frac{n^2}{2^{(\varepsilon-2\alpha)\sqrt{\lg n}}}
 \leq\frac{n^2}{2^{\alpha\sqrt{\lg n}}}. \qedhere
\]
\end{proof}

\begin{proof}[Proof of \Cref{thm:derandomized}]
The preprocessing uses $O(n^2)$ bits and is deterministic.  The query
algorithm is correct by \Cref{subsec:answering}, and it runs in time
$n^2/2^{\Omega(\sqrt{\lg n})}$ by \Cref{lem:compute-W}.
\end{proof}

\subsection{Encoding arbitrary \texorpdfstring{$k$}{k}-CNF functions}\label{sec-reduction-to-antitone}

We first reduce $2$-CNF functions to independent-set constraints of bipartite graphs.

\begin{lemma}\label{lem-2cnf-to-is}
Let $\varphi$ be a $2$-CNF formula on $n\geq1$ variables. One can construct in polynomial time a bipartite graph $G$ with $2n$ vertices on each side with the following property: given an encoding of $\mathsf{IS}_G$ with $T$ clauses, one can construct in polynomial time an encoding of the function defined by $\varphi$ with at most $T+4n$ clauses.
\end{lemma}
\begin{proof}
If $\varphi$ contains the empty clause, then it defines the constant-$\zero$ function, and the empty clause is an encoding of it; in this case, $G$ can be arbitrary. Otherwise, let $\varphi'$ be the $2$-CNF formula obtained from $\varphi$ by replacing every positive literal $x_i$ by $\overline{y_i}$, where $y_i$ is a fresh variable. Then, $\varphi'$ is an \emph{antitone} $2$-CNF formula on the $2n$ variables $x_1,\ldots,x_n,y_1,\ldots,y_n$, meaning that all of its literals are negative. Thus, each clause of $\varphi'$ is either a unit clause $\overline{a}$ or a clause $\overline{a}\lor\overline{b}$ with $a\neq b$. Let $\mathcal U$ be the set of unit clauses of $\varphi'$, so $|\mathcal U|\leq2n$, and let $F$ be the graph on these $2n$ variables with an edge $\{a,b\}$ for each clause $\overline{a}\lor\overline{b}$ of $\varphi'$. Then, $\varphi'$ is equivalent to $\mathsf{IS}_F\land\bigwedge_{C\in\mathcal U}C$.

Since $F$ need not be bipartite, let $G$ be the bipartite graph with a left vertex $a^L$ and a right vertex $a^R$ for each vertex $a$ of $F$, and with the edges $\{a^L,b^R\}$ and $\{b^L,a^R\}$ for each edge $\{a,b\}$ of $F$. A set $X$ of vertices is independent in $F$ if and only if $\{a^L\mid a\in X\}\cup\{a^R\mid a\in X\}$ is independent in $G$. Thus, given an encoding of $\mathsf{IS}_G$ with $T$ clauses, substituting the variable $a$ for both input variables $a^L$ and $a^R$ gives an encoding $\psi'$ of $\mathsf{IS}_F$ with $T$ clauses. Finally,
\[
 \psi:=\psi'\land\bigwedge_{C\in\mathcal U}C\land\bigwedge_{i\in[n]}(x_i\lor y_i)\land\bigwedge_{i\in[n]}(\overline{x_i}\lor\overline{y_i}),
\]
with $y_1,\ldots,y_n$ regarded as auxiliary variables, is an encoding of $\varphi$ with at most $T+4n$ clauses. Indeed, the last two conjunctions force $y_i=\overline{x_i}$ for every $i$, and under this substitution, $\varphi'$ is equivalent to $\varphi$. All of these steps take polynomial time.
\end{proof}

Next, we lift encodings of $k$-CNF functions to $(k+1)$-CNF functions. This costs a factor of $2n$ in the number of clauses, and hence preserves any improvement factor over $n^{k}$.

\begin{lemma} \label{lem-guarded-reduction}
Let $k\geq2$. Suppose that every $k$-CNF function on $n\geq1$ variables has an encoding with at most $f(n)$ clauses for some $f\colon\mathbb N\to\mathbb N$. Then, every $(k+1)$-CNF function on $n$ variables has an encoding with at most $2n\cdot f(n)$ clauses. Moreover, this lifting is constructive: if encodings of $k$-CNF functions with at most $f(n)$ clauses can be computed from $k$-CNF formulas in polynomial time, then encodings of $(k+1)$-CNF functions with at most $2n\cdot f(n)$ clauses can be computed from $(k+1)$-CNF formulas in polynomial time.
\end{lemma}
\begin{proof}
Let $\varphi$ be a $(k+1)$-CNF formula on the variables $x_1,\ldots,x_n$. If $\varphi$ contains the empty clause, then it defines the constant-$\zero$ function, and the empty clause is an encoding of it. This encoding has at most $2n\cdot f(n)$ clauses, since $f(n)\geq1$.

Otherwise, given a literal $\ell$, let $\varphi_\ell:=\{C\mid(\ell\lor C)\in\varphi\}$, which is a $k$-CNF formula on $x_1,\ldots,x_n$. Every clause of $\varphi$ is nonempty, so it has the form $\ell\lor C$ for some literal $\ell$, and then $C\in\varphi_\ell$. Conversely, $\ell\lor C$ is a clause of $\varphi$ for every $C\in\varphi_\ell$. Thus,
\[
 \varphi=\bigwedge_{i\in[n]}(x_i\lor\varphi_{x_i})\land\bigwedge_{i\in[n]}(\overline{x_i}\lor\varphi_{\overline{x_i}}),
\]
where $\ell\lor\varphi_\ell$ denotes the CNF formula $\bigwedge_{C\in\varphi_\ell}(\ell\lor C)$. By hypothesis, the function defined by each $\varphi_\ell$ has an encoding $\psi_\ell$ with at most $f(n)$ clauses, and we choose these encodings with pairwise disjoint auxiliary variables. Let
\[
 \psi:=\bigwedge_{i\in[n]}(x_i\lor\psi_{x_i})\land\bigwedge_{i\in[n]}(\overline{x_i}\lor\psi_{\overline{x_i}}).
\]
Then, $\psi$ has at most $2n\cdot f(n)$ clauses. It is not hard to see that $\psi$ is an encoding of the function defined by $\varphi$. Furthermore, if the encodings $\psi_\ell$ can be computed in polynomial time, then so can $\psi$.
\end{proof}

\noindent
We can now prove \Cref{thm-quasi}.
\thmquasi*
\begin{proof}
First, let $G$ be a bipartite graph with $s$ vertices on each side, and let $A\in\{\zero,\one\}^{s\times s}$ be its biadjacency matrix. Writing $u,v\in\{\zero,\one\}^s$ for the sets of vertices on the two sides, we have
\[
 \mathsf{IS}_G(u,v)=\one\quad\Longleftrightarrow\quad
 u^TAv=\zero.
\]
\Cref{thm:derandomized} therefore evaluates $\mathsf{IS}_G$ from
$m=O(s^2)$ advice bits in time
\[
 t=\frac{s^2}{2^{\Omega(\sqrt{\lg s})}}.
\]
At most $t$ distinct advice bits are read.
Applying \Cref{thm:robson-advice} with $b\leq t$ gives
\[
 \enc(\mathsf{IS}_G)=\widetilde O(t+\sqrt{mt})
 =\frac{s^2}{2^{\Omega(\sqrt{\lg s})}}.
\]

Now let $f$ be a $2$-CNF function on $n$ variables. \Cref{lem-2cnf-to-is} reduces $f$ to $\mathsf{IS}_G$ for a bipartite graph with $s=2n$ vertices on each side, at the cost of at most $4n$ additional clauses, which gives the claimed bound for $k=2$. The bound for each fixed $k \geq 3$ now follows from \Cref{lem-guarded-reduction}.
\end{proof}

\subsection{Encoding sparse graphs}\label{sec-sparse-graphs}

We first state Feder and Motwani's result about biclique partitions~\cite{feder-motwani}, which is essentially a restatement of a result of Nechiporuk~\cite{nechiporuk}. Given a graph $G$, a \emph{biclique partition} is a set $\{B_1,\dots,B_k\}$, where each $B_i$ is a complete bipartite subgraph of $G$, and $E(G) = \bigsqcup_{i=1}^k E(B_i)$. The \emph{weight} of such a partition is $\sum_{i=1}^k |V(B_i)|$. The \emph{edge density} of a graph with $n$ vertices and $m$ edges is $m/\binom{n}{2}$.

\begin{lemma}[Feder and Motwani~\cite{feder-motwani}] \label{lem-sparse-biclique-partition}
    Let $G = (V, E)$ be a graph with $n \ge 2$ vertices and edge density $\gamma > 0$. Then, $G$ has a biclique partition of weight $O\lrp{\frac{\gamma n^2 \lg(2/\gamma)}{\lg n}}$. Moreover, such a biclique partition can be constructed deterministically in time polynomial in $n$.
\end{lemma}

\noindent
Using this result, we can efficiently encode the independent set constraint for sparse graphs.

\begin{lemma} \label{lem-sparse-encoding}
    Let $G = (V, E)$ be a graph with $n \ge 2$ vertices and edge density at most $\gamma$, where $0 < \gamma \le 1/2$. Then, $\enc(\mathsf{IS}_G) = O\lrp{\frac{\gamma n^2 \lg(\gamma^{-1})}{\lg n}}$. Moreover, such an encoding can be constructed deterministically in time polynomial in $n$.
\end{lemma}
\begin{proof}
    Let $\gamma'$ be the edge density of $G$. If $\gamma' = 0$, then the empty formula is an encoding. Otherwise, by \Cref{lem-sparse-biclique-partition}, $G$ has a biclique partition $\mathcal{B} = \{B_1, \dots, B_k\}$ of weight $O\lrp{\frac{\gamma' n^2 \lg(2/\gamma')}{\lg n}}$, which can be constructed in polynomial time. Since the function $x \mapsto x \lg(2/x)$ is increasing on $(0, 1/2]$ and $\lg(2/\gamma) \le 2\lg(\gamma^{-1})$, this weight is $O\lrp{\frac{\gamma n^2 \lg(\gamma^{-1})}{\lg n}}$. We construct an encoding as follows. For each biclique $B_i = (X_i, Y_i)$, introduce an auxiliary variable $z_i$. Then, for each such biclique, include the following clauses in the encoding:
    \[
        \bigwedge_{x \in X_i} (\overline{x} \lor z_i) \land \bigwedge_{y \in Y_i} (\overline{z_i} \lor \overline{y}).
    \]
    The total number of clauses in the encoding is
    \[
        \sum_{i \in [k]} (|X_i| + |Y_i|) = O\lrp{\frac{\gamma n^2 \lg(\gamma^{-1})}{\lg n}}.
    \]

    Now, we prove that the encoding is correct. Let $X \subseteq V$ be an independent set. For each $i \in [k]$, set $z_i$ to true if and only if $X \cap X_i \neq \emptyset$. Then, every clause of the form $\overline{x} \lor z_i$ is satisfied. Also, if $z_i$ is true, then $X \cap Y_i = \emptyset$, since $X$ is independent and every vertex of $X_i$ is adjacent to every vertex of $Y_i$. Thus, every clause of the form $\overline{z_i} \lor \overline{y}$ is satisfied. Therefore, the encoding is satisfiable.

    On the other hand, suppose that $X \subseteq V$ is not independent. Let $x_1, x_2 \in X$ be such that $\{x_1,x_2\} \in E$. Then, $\{x_1,x_2\}$ is covered by some biclique $B_i$. Thus, our encoding either contains the pair of clauses $\overline{x_1} \lor z_i$ and $\overline{z_i} \lor \overline{x_2}$ or the pair of clauses $\overline{x_2} \lor z_i$ and $\overline{z_i} \lor \overline{x_1}$. In either case, the clauses imply the falsified clause $\overline{x_1} \lor \overline{x_2}$. Therefore, the encoding is unsatisfiable.
\end{proof}

\subsection{Encoding bipartite graphs with small \texorpdfstring{$\abip$}{alpha\_b}}\label{sec-small-abip}

Given a bipartite graph $G = (V_1 \sqcup V_2, E)$, its \emph{bipartite independence number}, denoted $\abip(G)$, is the largest $k$ such that $G$ has an independent set $S$ with $|S \cap V_1| = |S \cap V_2| = k$.

\subsubsection{Hash table lemma}\label{sec-hash}

Our encoding of the independent set constraint for graphs with small $\abip$, \Cref{thm-small-abip}, requires a hash table construction. As in \Cref{sec:non-uniform}, we formalize the hash table as a bipartite graph, whose left side represents the universe of elements that can be hashed and whose right side represents the slots of the table, and we use Teutsch's explicit expander (\Cref{lem:teutsch}). For our application, we need a uniform bound on the degree of vertices on the right side. We achieve this from \Cref{lem:teutsch} by splitting vertices on the right side whose degree is too high.

\begin{lemma} \label{lem-expander-peel}
    There is a function $n \mapsto d_n$ with $1 \le d_n = (\lg n)^{O(1)}$ such that the following holds. Let $L$ be a set of size $n \ge 1$, and let $r$ be an integer with $1 \le r \le n$. Then, there is a bipartite graph $(L \sqcup R, E)$ such that (i) $|R| \le 17r$, (ii) for every $S \subseteq L$ with $|S| \le r$, there is a matching that saturates $S$, and (iii) every vertex of $L$ has degree at most $d_n$, and every vertex of $R$ has degree at most $\ceil{n d_n / r}$. Moreover, this graph can be constructed deterministically in time polynomial in $n$.
\end{lemma}
\begin{proof}
    Let $k = \ceil{\lg r}$, so that $r \le 2^k < 2r$. If $2^k > n$, then we give each element of $L$ its own slot: let $R$ be a copy of $L$ and match each element to its copy. Then, $|R| = n < 2^k < 2r$, and all degrees are $1$, so the lemma holds.

    Otherwise, $n \ge 2^k$. By \Cref{lem:teutsch}, there is a bipartite graph $(L \sqcup R', E')$ with $|R'| < 2^{k+3} < 16r$ such that every $S \subseteq L$ with $|S| \le 2^k$ satisfies $|N(S)| \ge |S|$, and every vertex of $L$ has degree bounded by a fixed polynomial in $\lg n$. Let $d_n$ be the maximum of $1$ and this bound. By Hall's theorem~\cite{hall}, every $S \subseteq L$ with $|S| \le r \le 2^k$ has a matching that saturates $S$. Furthermore, $|E'| \le n d_n$.

    Let $\Delta := \ceil{n d_n / r}$. We construct $(L \sqcup R, E)$ from $(L \sqcup R', E')$ by doing the following operation on each $v \in R'$: split $v$ into $\ceil{d(v)/\Delta}$ vertices $v_1, \dots, v_{\ceil{d(v)/\Delta}}$ such that $d(v_i) \le \Delta$ for each $i$ and the sets $N(v_i)$ partition $N(v)$. We have
    \[
        |R| = \sum_{v \in R'} \ceil{d(v)/\Delta} \le |R'| + \frac{1}{\Delta} \sum_{v \in R'} d(v) \le |R'| + \frac{|E'|}{\Delta} < 16r + r = 17r.
    \]
    Every vertex of $L$ retains its degree, and every vertex of $R$ has degree at most $\Delta$ by construction. Furthermore, a matching that saturates $S$ in $(L \sqcup R', E')$ lifts to a matching that saturates $S$ in $(L \sqcup R, E)$, by replacing the right endpoint of each matching edge with the copy containing that edge; distinct right endpoints have distinct copies. Finally, by \Cref{lem:teutsch}, all $|E'| \le n d_n$ edges can be computed in time polynomial in $n$, and the splitting takes polynomial time.
\end{proof}

\subsubsection{The small \texorpdfstring{$\abip$}{alpha\_b} encoding}

The explicit hash table of \Cref{lem-expander-peel} has polylogarithmic left degree, so we state the encoding only in the range $\alpha \ge n / \lg n$, which suffices for our application. 

\begin{lemma} \label{thm-small-abip}
    Let $G = (V_1 \sqcup V_2, E)$ be a bipartite graph with $n$ vertices, and let $\alpha$ be an integer with $n / \lg n \le \alpha \le n/2$ and $\abip(G) \le \alpha$. Then, $\enc(\mathsf{IS}_G) = O\lrp{\frac{n \alpha \lg\lg n}{\lg n}}$. Moreover, given $G$ and $\alpha$, such an encoding can be constructed deterministically in time polynomial in $n$.
\end{lemma}
\begin{proof}
    For $n$ smaller than a sufficiently large absolute constant, we use one clause per edge, so assume that $n$ is sufficiently large. By \Cref{lem-expander-peel} with $L = V_1 \sqcup V_2$ and $r = \alpha$, there is a bipartite graph $H = (L \sqcup R, E')$ such that (i) $|R| \le 17\alpha$, (ii) for every $S \subseteq V_1 \sqcup V_2$ with $|S| \le \alpha$, there is a matching that saturates $S$, and (iii) every $i \in V_1 \sqcup V_2$ has degree at most $d_n$, and every $j \in R$ has degree at most $\ceil{n d_n / \alpha} \le d_n \lg n + 1$. Extend $H$ by adding a vertex $0$ to $L$ (we assume that $0 \notin V_1 \sqcup V_2$) such that $N_{H}(0) = R$. Let $d := \ceil{d_n \lg n} + 2$, so that $|N_{H}(j)| \le d$ for every $j \in R$. Let $t = \floor{\log_d \sqrt[3]{n}}$. Since $d = (\lg n)^{O(1)}$ and $n$ is sufficiently large, we have $t \ge 1$. Partition $R$ into parts $S_1, \dots, S_K$, where $K := \ceil{|R|/t}$, each of size at most $t$. For each $k \in [K]$, let $A_k  = \prod_{j \in S_k} N_{H}(j)$. Given $\vec{a} \in A_k$ and $j \in S_k$, we write $\vec{a}(j)$ to denote the element of $N_{H}(j)$ occupying the $j$-component of $\vec{a}$, and we write $\vec{a} \cap N_G(i) = \emptyset$ if $\vec{a}(j) \notin N_G(i)$ for every $j \in S_k$.

    We introduce the following auxiliary variables: $b$, $h_{i,j}$ for all $i \in V_1 \sqcup V_2 \sqcup \{0\}$ and $j \in N_{H}(i)$, and $z_{k,\vec{a}}$ for all $k \in [K]$ and $\vec{a} \in A_k$. Then, our encoding is as follows:
    \begin{align*}
        &\bigwedge_{i \in V_1} \lrp{\overline{x_i} \lor \overline{b} \lor \bigvee_{j \in N_{H}(i)} h_{i,j}} \land \bigwedge_{i \in V_2} \lrp{\overline{x_i} \lor b \lor \bigvee_{j \in N_{H}(i)} h_{i,j}} \land \tag{1} \\
        &\bigwedge_{k \in [K]} \amo\lrp{\left\{z_{k,\vec{a}} \mid \vec{a} \in A_k\right\}} \land \tag{2} \\
        &\bigwedge_{k \in [K]} \bigwedge_{j \in S_k} \bigwedge_{i \in N_{H}(j)} \lrp{\overline{h_{i,j}} \lor \bigvee_{\substack{\vec{a} \in A_k \\ \vec{a}(j) = i}} z_{k, \vec{a}}} \land \tag{3} \\
        &\bigwedge_{i \in V_1 \sqcup V_2} \bigwedge_{k \in [K]} \lrp{\overline{x_i} \lor \bigvee_{\substack{\vec{a} \in A_k \\ \vec{a} \cap N_G(i) = \emptyset}} z_{k, \vec{a}}}. \tag{4}
    \end{align*}

    First, we justify the correctness of the encoding. Given an assignment to the input variables, let $X$ be the set of vertices of $G$ corresponding to true input variables, so $\{x_i \mid i \in X\}$ is the set of true input variables. Suppose first that $X$ is an independent set in $G$. We either have $|X \cap V_1| \le \abip(G) \le \alpha$ or $|X \cap V_2| \le \alpha$, since otherwise $X$ contains an independent set with $\abip(G) + 1$ vertices on each side. If the former case holds, then set $b$ to true; otherwise, set $b$ to false. From now on, we assume that $|X \cap V_1| \le \alpha$, since the other case is symmetric. There is a matching in $H$ that saturates $X \cap V_1$. Given $i \in X \cap V_1$, let $h(i) \in N_{H}(i)$ be the vertex paired with $i$ according to the matching. For each $i \in V_1 \sqcup V_2$ and $j \in N_{H}(i)$, set $h_{i,j}$ to true if and only if $i \in X \cap V_1$ and $h(i) = j$. For each $j \in R$, set $h_{0,j}$ to true if and only if there is no $i \in X \cap V_1$ with $h(i) = j$. Then, for each $j \in R$, there is exactly one $i \in N_{H}(j)$ such that $h_{i,j}$ is true; call it $i_j$. For each $k \in [K]$, let $\vec{a}_k \in A_k$ be the tuple with $\vec{a}_k(j) = i_j$ for each $j \in S_k$, and set $z_{k,\vec{a}}$ to true if and only if $\vec{a} = \vec{a}_k$. We claim that with this assignment to the auxiliary variables, each clause is satisfied:
    \begin{itemize}
        \item If $x_i$ is true and $i \in V_1$, then $h_{i,h(i)}$ is true. Since we are assuming that $b$ is true, the clauses on line (1) are satisfied.
        \item The clauses on line (2) are satisfied, because for each $k \in [K]$, we set $z_{k,\vec{a}}$ to true for exactly one $\vec{a} \in A_k$.
        \item The clauses on line (3) are satisfied, because if $h_{i,j}$ is true for some $j \in S_k$, then $i = i_j = \vec{a}_k(j)$, and $z_{k,\vec{a}_k}$ is true.
        \item Suppose $x_i$ is true, and let $k \in [K]$. Then, $z_{k,\vec{a}_k}$ is true, and $\vec{a}_k(j) \in (X \cap V_1) \sqcup \{0\}$ for all $j \in S_k$. Since $X$ is an independent set and $i \in X$, we have $\vec{a}_k \cap N_G(i) = \emptyset$. Thus, the clauses on line (4) are satisfied.
    \end{itemize}

    Suppose next that $X$ is not an independent set in $G$. Then, there is some $i_1 \in X \cap V_1$ and $i_2 \in X \cap V_2$ such that $\{i_1,i_2\} \in E$. Then, by the clauses in line (1), we have $h_{i,j}$ true for some $i \in \{i_1,i_2\}$ and $j \in R$. Without loss of generality, $h_{i_1,j}$ is true. Let $k \in [K]$ be such that $j \in S_k$. Then, by the clauses in line (3), $z_{k,\vec{a}}$ is true for some $\vec{a} \in A_k$ with $\vec{a}(j) = i_1$. Also, by the clauses in line (4), $z_{k,\vec{a}'}$ is true for some $\vec{a}' \in A_k$ with $\vec{a}' \cap N_G(i_2) = \emptyset$. Since $i_1 \in N_G(i_2)$, we have $\vec{a} \neq \vec{a}'$, contradicting the clauses in line (2). This completes the proof that the encoding is correct.

    Now, we count the number of clauses in the encoding. Since $t \ge 1$, we have $t \ge \frac{1}{2}\log_d \sqrt[3]{n} = \frac{\lg n}{6 \lg d}$, and $\lg d = O(\lg \lg n)$. Recall also that $\alpha \ge n / \lg n$. We count the number of clauses as follows:
    \begin{itemize}
        \item There are $n = O\lrp{\frac{n \alpha \lg\lg n}{\lg n}}$ clauses on line (1).
        \item For each $k \in [K]$, we have
        \[
            |A_k| \le d^t \le \sqrt[3]{n},
        \]
        so by \Cref{prop:amo-bound}, the number of clauses on line (2) is
        \[
            K \cdot O\lrp{\sqrt[3]{n}} = O\lrp{\alpha \sqrt[3]{n}} = O\lrp{\frac{n \alpha \lg\lg n}{\lg n}},
        \]
        since $K \le |R| \le 17\alpha$.
        \item There is one clause on line (3) for each edge of $H$, so there are at most $n d_n + |R| \le n d_n + 17\alpha = O\lrp{\frac{n \alpha \lg\lg n}{\lg n}}$ clauses on line (3), since $d_n = (\lg n)^{O(1)}$ and $\alpha \ge n/\lg n$.
        \item The number of clauses on line (4) is
        \[
            n K \le n \lrp{\frac{17\alpha}{t} + 1} = O\lrp{\frac{n \alpha \lg d}{\lg n} + n} = O\lrp{\frac{n \alpha \lg\lg n}{\lg n}}.
        \]
    \end{itemize}
    Thus, the total number of clauses is $O\lrp{\frac{n \alpha \lg\lg n}{\lg n}}$, as desired. Finally, $H$ can be constructed in polynomial time by \Cref{lem-expander-peel}, and since $|A_k| \le \sqrt[3]{n}$ for each $k$, all clauses can be written in polynomial time.
\end{proof}

\subsection{Encoding arbitrary bipartite graphs}
\label{sec-constructive-graphs}

At this point of the argument, we decompose an arbitrary bipartite graph into components to which \Cref{lem-sparse-encoding,thm-small-abip} apply, by recursively splitting its biadjacency matrix into squares. Given a bipartite graph $G = (A \sqcup B, E)$, we identify each edge $\{a, b\}$ with $a \in A$ and $b \in B$ with the pair $(a,b) \in A \times B$. Given $U \subseteq A$ and $W \subseteq B$, we call $U \times W$ a \emph{rectangle} with sides $U$ and $W$, and a \emph{square} if $|U| = |W|$; we write $e(U,W) := |E \cap (U \times W)|$ and $G[U,W] := (U \sqcup W, E \cap (U \times W))$. The \emph{density} of $U \times W$ is
\[
    d(U,W) := \frac{e(U,W)}{|U| \cdot |W|},
\]
and $U \times W$ is \emph{empty} if $e(U,W) = 0$. Thus, $\abip(G[U,W])$ is the largest $k$ such that $U \times W$ contains an empty $k \times k$ square.

Deciding whether a bipartite graph has a large empty square is NP-complete~\cite{DBLP:books/fm/GareyJ79}. Instead, we use the following search, which finds a sparse square whenever a large empty square exists. We state it for binary matrices, which should be interpreted as biadjacency matrices, whose rows and columns are indexed by the two sides of a bipartite graph, so that a rectangle of the matrix is a rectangle $U \times W$, and its density is $d(U,W)$.

\begin{lemma}[Deterministic sparse-square search]
\label{lem-sparse-rectangle-search}
Given an $s \times s$ binary matrix, rational numbers $0 < \alpha \leq 1$ and  $0 < \delta < \nicefrac{1}{4}$, and an integer $u$ such that $1\leq u \leq \alpha s / 2$, with the property that the matrix contains an empty square $I \times J$ with $|I| = |J| \geq \alpha s$, there is a deterministic algorithm that finds a $u \times u$ square of density at most $4\delta$ in time
\[
 s^{O(1)}
 \exp\!\left(O\!\left(\delta^{-1}\lg^2\frac1{\alpha\delta}\right)\right).
\]
\end{lemma}

Our proof uses the following bounds of Alon, Feige, Wigderson, and Zuckerman on the probability that a random walk stays inside a set; we state the special case in which all eigenvalues are nonnegative.

\begin{lemma}[Alon--Feige--Wigderson--Zuckerman~{\cite[Proposition~2]{DBLP:journals/cc/AlonFWZ95}}] \label{lem-afwz}
    Let $P$ be the transition matrix of the random walk on a regular multigraph on $s$ vertices, and suppose that $P$ is positive semidefinite and every eigenvector of $P$ orthogonal to the all-ones vector $\mathbf{1}$ has eigenvalue at most $\lambda$. Let $T$ be a set of $\mu \cdot s$ vertices, and let $w_0, \ldots, w_L$ be a random walk with transition matrix $P$ and a uniformly random starting vertex. Then,
    \[
        \Pr[w_0, \ldots, w_L \in T] \le \mu(\mu + \lambda)^L,
    \]
    and if $L$ is even, then also $\Pr[w_0, \ldots, w_L \in T] \ge \mu^{L+1}$.
\end{lemma}
\begin{proof}[Proof of \Cref{lem-sparse-rectangle-search}]
We prove the claim assuming $s=m^2$ is a perfect square, since otherwise the same argument can be made by padding the matrix with $1$s until its dimension is the next perfect square.\footnote{With $s' := \lceil \sqrt{s}\,\rceil^2 < 4s$, replacing $\alpha$ by
$\alpha s/s' > \alpha/4$ preserves the hypotheses and changes the running time only by
constant factors.}
The proof proceeds conceptually by considering a random walk over possible squares, and showing that it succeeds in finding a sparse square, without taking too many steps, with positive probability. The deterministic algorithm is based on enumerating all short random walks of a suitable explicit expander.

We label the $s$ rows of the input matrix $M$ by $\mathbb Z_m^2$, and use them as the vertex set of an $8$-regular expander multigraph on $s$ vertices, constructed as per Margulis--Gabber--Galil~\cite[Construction~8.1]{HooryLinialWigderson2006Expanders}, meaning that each vertex $(x,y) \in \mathbb Z_m^2$ has an edge to the following vertices:
\[
\begin{aligned}
 &(x\pm2y,y), &&(x\pm(2y+1),y),\\
 &(x,y\pm2x), &&(x,y\pm(2x+1)),
\end{aligned}
\]
where all arithmetic is modulo $m$. Let $A$ be the adjacency matrix of this graph, and $Q_0 :=A/8$ its transition matrix.
By~\cite[Theorem~8.2]{HooryLinialWigderson2006Expanders}, the constant vector $\mathbf{1} \in \mathbb{R}^s$ is an eigenvector of $Q_0$ with eigenvalue $1$, and every eigenvector in the set $\mathbf{1}^\perp := \{ v \in \mathbb{R}^s \mid \langle v, \mathbf{1}\rangle = 0\}$ has an eigenvalue with absolute value at most $5\sqrt2/8 < 0.89$.

We then consider the lazy random walk of transition matrix
\[
 Q:=\frac{\mathrm{Id}_s+Q_0}{2},
\]
meaning that it stays at its current row with probability $1/2$ and otherwise takes a step according to $Q_0$.
Note that $Q$ is symmetric, and since each eigenvalue $\theta$ of $Q_0$ corresponds to an eigenvalue $(1+\theta)/2 \geq 0$ of $Q$, we have that $Q$ is positive semidefinite, and its eigenvalues on $\mathbf1^\perp$ are at most
\[
 \frac{1+5\sqrt2/8}{2}<\frac{19}{20}.
\]
Note that taking a step according to $Q$ can be seen as choosing uniformly from $16$ possibilities ($8$ directions and $8$ choices representing staying at the given row). 

Instead of considering $Q$ directly, we use a power of $Q$ in order to further reduce the magnitude of the eigenvalues.
Let $b := \ceil{\lg(2/(\alpha\delta))}$, and $h := 20b$. We consider $P := Q^h$, which corresponds to taking $h$ steps according to $Q$. Therefore, each step according to $P$ can be seen as choosing uniformly from $16^h$ possibilities.
The eigenvalues of $P$ on $\mathbf1^\perp$ lie in $[0,\lambda]$, where
\[
 \lambda\leq(19/20)^{20b}\leq e^{-b}\leq2^{-b}
          \leq\alpha\delta/2.
\]
Let $L := 2\ceil{b/\delta}$ be the length of the walks we consider. The algorithm enumerates all walks $W := (w_0, w_1, \ldots, w_L)$ with transition matrix $P$, over all starting rows and all step choices. The number of such walks is at most
\[
    s \cdot (16^h)^L = s \exp\left( O\left(\frac{b^2}{\delta}\right)\right) = s \cdot \exp\left(O\left(\delta^{-1} \lg^2 \frac{1}{\alpha\delta}\right)\right).
\]

For each walk $W$, we attempt to get a sparse square in the original matrix $M$ as follows:
\begin{enumerate}
    \item Let $C \subseteq [s]$ be the columns of $M$ whose entries are zero in every row visited by $W$.
    \item Let $R \subseteq [s]$ be the set of rows  $r$ such that $M$ has at most $4\delta|C|$ ones on row $r$ in the columns $C$.
    \item If $|R|<u$ or $|C|<u$, continue to the next walk.
    \item Otherwise, choose any $u$ rows $R'\subseteq R$ and the $u$ columns $C'\subseteq C$ with the fewest ones in those rows, and return $R'\times C'$.
\end{enumerate}
Writing $e(X,Y)$ for the number of ones in $M[X\times Y]$, the choice of $C'$ gives
\[
 e(R',C')\leq\frac{u}{|C|}e(R',C)
              \leq4\delta u^2.
\]
Thus every returned square has the required density. If no walk succeeds, the algorithm reports failure. Note immediately that each of the four steps above takes time polynomial in $s$, and thus the final runtime is upper bounded by the number of walks times a polynomial in $s$, satisfying the desired bound.

We now prove that some walk succeeds whenever the promised empty square $I\times J$ exists. Put $\beta:=|I|/s\geq\alpha$, and consider a random walk $w_0,\ldots,w_L$ with transition matrix $P$ and a uniformly chosen starting row. Let $\mathcal E$ be the event that all $L+1$ visited rows lie in $I$. Since $L$ is even, \Cref{lem-afwz} applied to $T := I$ gives
\[
 \Pr[\mathcal E]\geq\beta^{L+1}.
\]

Call a column bad if it has at least $\delta|I|$ ones in the rows of $I$. For a bad column $j$, let $S\subseteq I$ be the rows where its entry is zero. Then $|S|/s\leq(1-\delta)\beta$. Conditional on $\mathcal E$, the column belongs to $C$ precisely when the walk stays entirely in $S$. By \Cref{lem-afwz} applied to $T := S$, we have
\[
 \Pr[w_0,\ldots,w_L\in S]
 \leq\frac{|S|}{s}\left(\frac{|S|}{s}+\lambda\right)^L
 \leq\beta\bigl((1-\delta)\beta+\lambda\bigr)^L.
\]
Dividing by the lower bound on $\Pr[\mathcal E]$ and using $\lambda/\beta\leq\delta/2$, we obtain
\[
 \Pr[j\in C\mid\mathcal E]
 \leq(1-\delta+\lambda/\beta)^L
 \leq e^{-\delta L/2}
 \leq e^{-b}
 \leq\alpha\delta.
\]
There are at most $s$ bad columns, so, conditional on $\mathcal E$, the expected number of bad columns in $C$ is at most $\alpha\delta s$. Hence some walk entirely inside $I$ leaves at most that many bad columns in $C$.

For this walk, $J\subseteq C$ because $I\times J$ is empty, so $|C|\geq\alpha s$. Each column that is not bad contributes fewer than $\delta|I|$ ones in the rows of $I$, and each bad column contributes at most $|I|$. Therefore,
\[
 e(I,C)\leq\delta|I||C|+\alpha\delta s|I|
          \leq2\delta|I||C|.
\]
The average number of ones in $C$ over the rows of $I$ is thus at most $2\delta|C|$. At least half those rows have at most $4\delta|C|$ ones, so $|R|\geq|I|/2\geq u$. We also have $|C|\geq\alpha s\geq2u$, and the algorithm succeeds.
\end{proof}

We now state the graph decomposition. It can be seen as a kind of regularity lemma~\cite{szem}: every square of the decomposition is pseudorandom in the sense that it contains no large empty square, and all remaining edges form a sparse graph.

\begin{lemma}[Sparse--pseudorandom decomposition] \label{lem-decomp}
    Let $G = (A \sqcup B, E)$ be a bipartite graph with $|A| = |B| = N$. Then, there is an integer $q = \Theta\lrp{\frac{\lg N}{(\lg\lg N)^2}}$ such that $A \times B$ can be partitioned into squares $U_1 \times W_1, \dots, U_p \times W_p$ and a set $S \subseteq A \times B$ such that:
    \begin{enumerate}[label=(\roman*)]
        \item every square has side length $m_i := |U_i| = |W_i| \ge \sqrt{N}$;
        \item for every $i \in [p]$, we have $\abip(G[U_i,W_i]) < 2\floor{m_i/q}$;
        \item $|E \cap S| = O(N^2/q)$.
    \end{enumerate}
    Moreover, $q$ and such a partition can be computed deterministically in time polynomial in $N$.
\end{lemma}
\begin{proof}
    If $N$ is smaller than a sufficiently large absolute constant, we take $q := 1$, $p := 0$, and $S := A \times B$. Otherwise, put
    \[
        t := \floor{\lg N}, \qquad \ell := \ceil{\lg t}, \qquad q := \floor{\frac{t}{64 \ell^2}},
    \]
    so that $q = \Theta\lrp{\frac{\lg N}{(\lg\lg N)^2}}$, and assume that $N$ is large enough that $q \ge 12$, $\sqrt{N} \ge 2q$, and $\ell \ge 6$. We describe a recursive procedure $\textsc{Split}(U,W)$, which takes a square $U \times W$ with $U \subseteq A$, $W \subseteq B$, and $|U| = |W| = m$, and partitions it into squares of the form required by (i) and (ii) and cells that are added to $S$:
    \begin{enumerate}[label=(\alph*)]
        \item If $m < \sqrt{N}$, add $U \times W$ to $S$. We call $U \times W$ a \emph{small square}.
        \item Otherwise, let $u := \floor{m / q}$, and write $m = qu + r$ with $0 \le r < q$. Since $m \ge 2q$, we have $u \ge m/(2q) \ge 1$. Apply \Cref{lem-sparse-rectangle-search} to the biadjacency matrix of $G[U,W]$ with $\alpha := 2u/m$, $\delta := 1/q$, and target side length $u$; its hypotheses hold since $\alpha \le 2/q \le 1$, $\delta < 1/4$, and $u = \alpha m / 2$.
        \begin{enumerate}[label=(b\arabic*)]
            \item If the search reports failure, then $U \times W$ contains no empty square of side length at least $\alpha m = 2u$, so $\abip(G[U,W]) < 2u$. Output $U \times W$ as one of the squares $U_i \times W_i$.
            \item Otherwise, the search returns $U' \subseteq U$ and $W' \subseteq W$ with $|U'| = |W'| = u$ and $e(U', W') \le 4u^2/q$. Add $U' \times W'$ to $S$; we call it a \emph{sparse square}. Since $m - u = (q-1)u + r$, we can partition $U \setminus U'$ into sets $U'_1, \dots, U'_{q-1}$ of size $u$ and a set $U'_0$ of size $r$, and similarly partition $W \setminus W'$ into sets $W'_1, \dots, W'_{q-1}$ of size $u$ and a set $W'_0$ of size $r$. Add $U' \times W'_0$ and $U'_0 \times W'$ to $S$; we call them \emph{slivers}. Finally, call $\textsc{Split}(U \setminus U', W \setminus W')$, and call $\textsc{Split}(U', W'_j)$ and $\textsc{Split}(U'_j, W')$ for every $j \in [q-1]$. We call $\textsc{Split}(U \setminus U', W \setminus W')$ the \emph{large child} of this call, and the \emph{large-child branch} of a call is the sequence of calls obtained by repeatedly passing to the large child.
        \end{enumerate}
    \end{enumerate}
    In case (b2), the sparse square, the two slivers, and the squares of the recursive calls partition $U \times W$. Thus, $\textsc{Split}(A,B)$ produces a partition of $A \times B$ into squares $U_i \times W_i$ and a set $S$, which consists of sparse squares, slivers, and small squares. Properties (i) and (ii) hold by construction, since squares are only output in case (b1). It remains to prove (iii) and to bound the running time.

    We first bound the running time. Every recursive call is on a square of side length $m - u$ or $u$, both of which are positive and smaller than $m$, so the recursion terminates. The calls that do not recurse are on pairwise disjoint nonempty squares, so there are at most $N^2$ of them, and every call that recurses makes at least two recursive calls, so there are at most $2N^2$ calls in total. In each call, $\delta^{-1} = q$ and $\alpha\delta = 2u/(mq) \ge 1/q^2$, so by \Cref{lem-sparse-rectangle-search} the search takes time
    \[
        m^{O(1)} \exp\lrp{O\lrp{q \ln^2 q}} = N^{O(1)},
    \]
    since $\ln q \le \ln t \le \ell$ and therefore $q \ln^2 q \le t / 64 \le \lg N$. Thus, the decomposition can be computed in time polynomial in $N$.

    Now, we prove (iii) by bounding the number of edges in the sparse squares, slivers, and small squares:
    \begin{itemize}
        \item Every sparse square $U' \times W'$ has $e(U',W') \le 4u^2/q = \frac{4}{q}|U'||W'|$. Since the sparse squares are pairwise disjoint, they contain at most $4N^2/q$ edges in total.
        \item Consider a call in case (b2) on a square of side length $m$. Its two slivers have $2ur < 2qu \le 2m$ cells, and its sparse square has $u^2 \ge m^2/(4q^2)$ cells. Hence, the slivers have at most
        \[
            2m \le \frac{8q^2 u^2}{m} \le \frac{8q^2 u^2}{\sqrt{N}}
        \]
        cells. Since the sparse squares are pairwise disjoint, summing over all calls in case (b2) shows that the slivers have at most $8q^2 N^2/\sqrt{N} = 8q^2 N^{3/2} = O(N^2/q)$ cells in total, since $q = O(\lg N)$.
        \item To bound the small squares, group consecutive calls along the large-child branch. A group starts with side length $m\ge2\sqrt{N}$ and follows this branch until its side length is at most $m/2$ or a search reports failure. Its children are the calls on squares of side length $u$ produced along the branch and the final large-child call, if one exists. Stop grouping at squares of side length below $2\sqrt{N}$. Every original small square lies within one of these squares.

        For every call in a group, its side length $a$ satisfies $m/2<a\le m$ and $a\ge\sqrt{N}\ge2q$, so $u=\floor{a/q}\ge m/(4q)$. If the group reaches side length at most $m/2$, its sparse squares satisfy $\sum u\ge m/2$, hence $\sum u^2\ge m^2/(8q)$. Otherwise, the search failure outputs a square of area at least $m^2/4$. Thus, the children have total area at most $\rho m^2$, where $\rho:=1-1/(8q)$. Every child has side length at least $m/(4q)$; the final large child has side length at least $m/2-m/q$.

        Put $d:=\floor{\log_{4q}(\sqrt{N}/2)}$. At depth $j$ of the grouped recursion, every square has side length at least $N/(4q)^j$, so no square of side length below $2\sqrt{N}$ occurs at depth at most $d$. All small squares therefore lie within the squares at depth $d$, whose total area is at most $N^2\rho^d$. Since $\ln(4q)\le\ln4+(\ln2)\ell\le\ell$, we have
        \[
            \frac{d}{8q}
            \ge \frac{\ln N}{16q\ln(4q)}-\frac{1}{4q}
            \ge 4(\ln2)\ell-\frac{1}{4q}
            \ge 2\ell.
        \]
        Therefore, the small squares have at most $N^2\rho^d\le N^2e^{-2\ell}\le N^2/t\le N^2/q$ cells in total.
    \end{itemize}
    Therefore, $|E \cap S| \le 4N^2/q + 8q^2 N^{3/2} + N^2/q = O(N^2/q)$.
\end{proof}

We can now prove our upper bound on encoding the independent set constraint for arbitrary bipartite graphs.

\begin{lemma}\label{thm-graph-polytime}
    Let $G = (A \sqcup B, E)$ be a bipartite graph with $|A| = |B| = N$. Then, $\enc(\mathsf{IS}_G) = O\lrp{\frac{N^2 (\lg\lg N)^3}{\lg^2 N}}$. Moreover, such an encoding can be constructed deterministically in time polynomial in $N$.
\end{lemma}
\begin{proof}
    For $N$ smaller than a sufficiently large absolute constant, we use one clause per edge, so assume that $N$ is sufficiently large.

    Apply \Cref{lem-decomp} to get a partition of $A \times B$ into squares $U_1 \times W_1, \dots, U_p \times W_p$ and a set $S$ of the stated form. We decompose $G$ into subgraphs whose edge sets partition $E$:
    \begin{itemize}
        \item For each $i \in [p]$, let $G_i := G[U_i, W_i]$ be the subgraph consisting of the edges in the square $U_i \times W_i$.
        \item Let $G_0 := (A \sqcup B, E \cap S)$ be the subgraph consisting of the edges in $S$, that is, the edges in sparse squares, slivers, and small squares.
    \end{itemize}
    A set of vertices is independent in $G$ if and only if it is independent in $G_0$ and its intersection with $U_i \sqcup W_i$ is independent in $G_i$ for every $i \in [p]$. Thus, the conjunction of encodings of $\mathsf{IS}_{G_0}$ and $\mathsf{IS}_{G_1}, \dots, \mathsf{IS}_{G_p}$, with pairwise disjoint auxiliary variables, is an encoding of $\mathsf{IS}_G$. We now show that these encodings use $O\lrp{\frac{N^2 (\lg\lg N)^3}{\lg^2 N}}$ clauses in total.

    For each $i \in [p]$, the graph $G_i$ is bipartite with $2m_i$ vertices, where $m_i \ge \sqrt{N}$. Let $u_i := \floor{m_i/q}$ and $\alpha_i := 2u_i$, so $\abip(G_i) < \alpha_i$ by \Cref{lem-decomp}. We check the remaining hypotheses of \Cref{thm-small-abip} with $n := 2m_i$ and $\alpha := \alpha_i$. Since $N$ is sufficiently large and $q = \Theta\lrp{\frac{\lg N}{(\lg\lg N)^2}}$, we have $2 \le q \le \frac{\lg N}{64}$ and $m_i \ge \sqrt{N} \ge 2q$. Thus, $\alpha_i \ge m_i/q$, and since $q \le \frac{\lg N}{64} \le \frac{\lg(2m_i)}{32}$, this gives $\alpha_i \ge \frac{2m_i}{\lg(2m_i)}$. Furthermore, $\alpha_i \le 2m_i/q \le m_i$. Thus, by \Cref{thm-small-abip}, $\mathsf{IS}_{G_i}$ can be encoded using
    \[
        O\lrp{\frac{m_i u_i \lg\lg m_i}{\lg m_i}} = O\lrp{\frac{m_i^2 \lg\lg N}{q \lg N}}
    \]
    clauses, since $u_i \le m_i / q$ and $\lg m_i \ge \frac{1}{2} \lg N$. Since the squares are pairwise disjoint, $\sum_{i \in [p]} m_i^2 \le N^2$, so the total number of clauses over all $i \in [p]$ is $O\lrp{\frac{N^2 \lg\lg N}{q \lg N}} = O\lrp{\frac{N^2 (\lg\lg N)^3}{\lg^2 N}}$, since $q = \Theta\lrp{\frac{\lg N}{(\lg\lg N)^2}}$.

    By \Cref{lem-decomp}, the graph $G_0$ has $2N$ vertices and at most $CN^2/q$ edges for some absolute constant $C$, so its edge density is at most
    \[
        \frac{CN^2}{q\binom{2N}{2}} \le \frac{C}{q} =: \gamma.
    \]
    Since $N$ is sufficiently large, we may assume that $0 < \gamma \le 1/2$. Therefore, by \Cref{lem-sparse-encoding}, we can encode $\mathsf{IS}_{G_0}$ using
    \[
        O\lrp{\frac{\gamma N^2 \lg(\gamma^{-1})}{\lg N}} = O\lrp{\frac{N^2 \lg q}{q \lg N}} = O\lrp{\frac{N^2 (\lg\lg N)^3}{\lg^2 N}}
    \]
    clauses, again since $q = \Theta\lrp{\frac{\lg N}{(\lg\lg N)^2}}$.

    Finally, the decomposition can be computed in polynomial time by \Cref{lem-decomp}, there are at most $N^2$ squares, and each encoding can be constructed in polynomial time by \Cref{thm-small-abip,lem-sparse-encoding}.
\end{proof}

\thmpolytime*
\begin{proof}
For $k=2$, apply \Cref{lem-2cnf-to-is} and encode $\mathsf{IS}_G$ by \Cref{thm-graph-polytime} with $N=2n$. For $k\geq3$, applying \Cref{lem-guarded-reduction} $k-2$ times gives $O_k(n^k(\lg\lg n)^3/\lg^2 n)$ clauses. Since each application is constructive, for fixed $k$ the encoding can be computed in polynomial time.
\end{proof}

\subsection{Counting lower bounds}\label{sec-kcnf-lb}

A $k$-CNF function has a width-$k$ encoding with $O_k(n^k/\lg n)$ clauses~\cite{ep-hypergraph,stoc}. The following counting argument shows that unbounded clause width is necessary to improve this bound, as in \Cref{thm-quasi,thm-polytime}.

\begin{proposition}\label{prop-kcnf-bounded-width}
For every fixed $k\ge2$ and $w\ge k$, there are $k$-\textrm{CNF} functions on $n$ inputs whose width-$w$ encodings require $\Omega_{k,w}(n^k/\lg n)$ clauses.
\end{proposition}
\begin{proof}
For each collection $\mathcal H\subseteq\binom{[n]}k$, consider the antitone $k$-CNF
\[
    f_{\mathcal H}(\vec x)=\bigwedge_{S\in\mathcal H}\bigvee_{i\in S}\overline{x_i}.
\]
On an assignment selecting exactly the elements of $S\in\binom{[n]}k$, this function is false exactly when $S\in\mathcal H$. Hence, these collections give $2^{\binom nk}$ distinct functions.

Let $q$ be the minimum integer such that these functions all have width-$w$ encodings with at most $q$ clauses. Since $w\ge k$, their original CNFs give $q\le\binom nk$. An encoding with at most $q$ clauses of width at most $w$ uses at most $wq$ occurring auxiliary variables. The number of possible clauses over $n+wq$ variables is $O_w((n+q)^w)$, so comparing the number of functions with the number of encodings gives
\[
    2^{\binom nk}\le2^{O_w(q\lg(n+q))}
    =2^{O_{k,w}(q\lg n)}.
\]
Thus $q=\Omega_{k,w}(n^k/\lg n)$.
\end{proof}

To run the same counting argument without a width bound, we need another way to bound the number of auxiliary variables. The following proposition shows that a minimal encoding has fewer auxiliary variables than clauses.
\begin{proposition}\label{prop:auxiliaries}
    Every boolean function $f$ with $\enc(f) \ge 1$ has an encoding $(\varphi, X, Y)$ such that $|\varphi| = \enc(f)$ and $|Y| <\enc(f)$.
\end{proposition}
Its proof relies on a graph-theoretic fact, similar in spirit to the matching argument of Aharoni and Linial for minimal unsatisfiable formulas~\cite{aharoniMinimalNontwocolorableHypergraphs1986b}.

\begin{lemma}\label{lemma:graph-theoretic}
    Let $G := (U \sqcup V, E)$ be a bipartite graph with $|V| \geq |U| > 0$ and such that all vertices in $V$ have degree at least $1$. Then, there is a nonempty subset $S \subseteq V$ for which $G$ contains a perfect matching between $S$ and $N_G(S)$.
\end{lemma}
\begin{proof}
    The proof is by induction on $|V|$. If $|V| = 1$ the only option is $|U| = 1$ and $E = \{\{u, v\}\}$ for $u \in U$ and $v \in V$, so the statement holds trivially.  
    Now, for the inductive case,
    let us say a matching $M$ is $V$-saturating if every vertex $v \in V$ belongs to some edge in $M$. Note that a $V$-saturating matching is a perfect matching between $V$ and $N_G(V) = U$. Thus, if a $V$-saturating matching exists we take $S := V$ and complete the proof.
    On the other hand, if no such $V$-saturating matching exists, then by Hall's condition~\cite{hall}, there exists a nonempty subset $W \subseteq V$ such that $|N_G(W)| < |W|$. We have $|W| \ge 2$, because otherwise $|N_G(W)| = 0$, contradicting our assumption that $V$ has no isolated vertices. Let $W' = W \setminus \{w\}$ for an arbitrary $w \in W$. Then, $W'$ is nonempty and $0 < |N_G(W')| \le |W'|$. Therefore, the induced bipartite graph $G' := G[N_G(W'), W']$ satisfies the inductive hypothesis, so there is a nonempty subset $S \subseteq W'$ for which $G'$ contains a perfect matching between $S$ and $N_{G'}(S)$. Since $N_{G'}(S) = N_G(S)$, the set $S \subseteq V$ satisfies the conclusion of the lemma.
\end{proof}

\noindent
Now, to prove~\Cref{prop:auxiliaries}, we need an elementary definition and a lemma.

\begin{definition}[Autarky~\cite{monienSolvingSatisfiabilityLess1985}]
    Given a \emph{CNF} formula $\varphi$ over a set of variables $X$, an autarky is a partial assignment $\alpha \colon X\to \{\zero, \one\}$ such that for any clause $C \in \varphi$ containing a variable in the domain of $\alpha$, we have $\alpha \models C$.
\end{definition}

\begin{lemma}[\cite{monienSolvingSatisfiabilityLess1985}, cf.~{\cite[Thm.~4]{kieslTruthAssignmentsConditional2019}}]\label{lemma:autarky}
    Let $\varphi$ be a \emph{CNF} formula and $\alpha$ an autarky for $\varphi$.
    Then, $\mathsf{SAT}(\varphi) \iff \mathsf{SAT}(\varphi_{\downharpoonright \alpha})$.\footnote{For a formula $F$ and an assignment $\tau$, notation $F_{\downharpoonright \tau}$ denotes the formula obtained by eliminating from $F$ all clauses already satisfied by $\tau$, and removing from each remaining clause the literals falsified by $\tau$.}
\end{lemma}

\noindent
We are now ready to prove~\Cref{prop:auxiliaries}.

\begin{proof}[Proof of~\Cref{prop:auxiliaries}]
    Let $(\varphi, X, Y)$ be an encoding for $f$ with $|\varphi| = \enc(f) \ge 1$. Let $Y' \subseteq Y$ be the subset of auxiliary variables that appear in $\varphi$ (with any polarity). Naturally, $(\varphi, X, Y')$ is also an encoding for $f$.
  Assume, expecting a contradiction, that $|Y'| \geq \enc(f)$. 
    Then, construct a bipartite graph $G_\varphi := (\varphi \sqcup Y', E)$ where there is an edge between an auxiliary variable $y \in Y'$ and a clause $C \in \varphi$ if and only if $y \in C$ or $\overline{y} \in C$. Now, observe that $G_{\varphi}$ satisfies the conditions of~\Cref{lemma:graph-theoretic}, so there exists a nonempty set $Y^\star \subseteq Y'$ and a set $\varphi^\star := N_{G_\varphi}(Y^\star)$ such that there is a perfect matching $M$ between $Y^\star$ and $\varphi^\star$. 
    Let $\alpha\colon Y^\star \to \{\zero, \one\}$ be the assignment defined by 
    \[
    \alpha(y)  = \begin{cases}
        \one & \text{if } y \in M(y)\\
        \zero & \text{if } \overline{y} \in M(y),
    \end{cases}
    \]
    and observe that $\alpha$ is an autarky for $\varphi$. Let $\tau \colon X \to \{\zero,\one\}$ be an arbitrary assignment to the input variables. Note that $\alpha$ and $\tau$ have disjoint domains, so $\alpha$ is still an autarky for $\varphi_{\downharpoonright\tau}$, and by \Cref{lemma:autarky}, we have $\textsf{SAT}(\varphi_{\downharpoonright\tau}) \iff \textsf{SAT}((\varphi_{\downharpoonright\tau})_{\downharpoonright\alpha})$. Since $(\varphi_{\downharpoonright\tau})_{\downharpoonright\alpha} = (\varphi_{\downharpoonright\alpha})_{\downharpoonright\tau}$, it follows that $(\varphi_{\downharpoonright\alpha}, X, Y \setminus Y^\star)$ is an encoding of $f$, and since $|\varphi_{\downharpoonright\alpha}| = |\varphi| - |\varphi^\star| < |\varphi|$, we conclude that $|\varphi_{\downharpoonright\alpha}| < \enc(f)$, a contradiction.
\end{proof}

\noindent
We now state the following counting lemma, which we use to prove all of our lower bounds for $\enc$.

\begin{lemma}\label{lem-counting}
    For all integers $n\geq1$ and $q\geq1$, at most $3^{(n+q)q}+1$ functions $f\colon\{\zero,\one\}^n\to\{\zero,\one\}$ satisfy $\enc(f)\leq q$.
\end{lemma}
\begin{proof}
    The constant-$\one$ function is the only $f$ with $\enc(f)=0$. If $1\leq\enc(f)\leq q$, then by~\Cref{prop:auxiliaries}, $f$ has an encoding with $\enc(f)$ clauses and fewer than $q$ auxiliary variables. This encoding has no tautological clauses, since deleting one would give a smaller encoding. Rename its auxiliary variables from a fixed set of $q$ names, and pad it to exactly $q$ clauses by repeating a clause. The result is a sequence of $q$ non-tautological clauses over $n+q$ fixed variables, and it determines $f$. Each such clause is determined by deciding, for each variable, whether it appears positively, appears negatively, or is absent, so there are at most $3^{n+q}$ of them. Hence at most $3^{(n+q)q}$ functions satisfy $1\leq\enc(f)\leq q$.
\end{proof}

Without a width bound, the counting argument, now based on \Cref{lem-counting}, gives a much weaker lower bound.

\begin{proposition}\label{prop-kcnf-unbounded-width}
For every fixed $k\ge2$, there are $k$-CNF functions on $n$ inputs with $\enc(f)=\Omega_k(n^{k/2})$.
\end{proposition}
\begin{proof}
For $n\ge k$, let $q:=\max_{\mathcal H}\enc(f_{\mathcal H})\ge1$, where the $f_{\mathcal H}$ are the $2^{\binom nk}$ functions from the proof of \Cref{prop-kcnf-bounded-width}. By \Cref{lem-counting}, $2^{\binom nk}\le3^{(n+q)q}+1$, hence
\[
    q\ge\sqrt{\frac{n^2}{4}+\log_3(2^{\binom nk}-1)}-\frac n2
    =\Omega_k(n^{k/2}). \qedhere
\]
\end{proof}

Together with \Cref{cor:random-kcnf}, \Cref{prop-kcnf-unbounded-width} shows that almost all $k$-CNF functions satisfy $\enc(f)=\widetilde\Theta(n^{k/2})$.\footnote{As usual, it is easy to see that the counting lower bounds above apply not only to some function in the class, but to a $1-o(1)$ fraction of them.} Thus, any better worst-case lower bound must come from functions that are far from random. Even for independent-set constraints ($k=2$), a large gap remains: \Cref{prop-kcnf-unbounded-width} gives only $\Omega(n)$, while the best upper bound we prove is $n^2/2^{\Omega(\sqrt{\lg n})}$ (\Cref{thm-quasi}). Closing this gap is a fascinating open problem. We expect the upper bound to be closer to the truth, but we currently lack a technique for proving superlinear lower bounds for $\enc(\mathsf{IS}_G)$.

\section{Arbitrary and Sparsely Supported Functions}\label{sec:arbitrary}

In this section, we prove \Cref{thm:sparse}, which we restate for convenience.
\thmsparse*

We give some extra consideration to the case of densely supported functions, establishing the following explicit
constant bounds:
\begin{proposition}\label{prop:dense}
    Let $\enc(n)$ be the maximum of $\enc(f)$ over all functions $f\colon \{\zero, \one\}^n \to \{\zero, \one\}$, and let $c := 1/\sqrt{\lg 3} \approx 0.794$. Then, we have $(c - o(1))\sqrt{2^n} \leq \enc(n) \leq 
    (2\sqrt{2} + o(1))\sqrt{2^n}$.
\end{proposition}
Establishing an exact constant for the upper and 
lower bounds in this model remains an open question. We expect that the upper bound is tight.

Bounds of the same order, often with exact constants, are classical in related models of computation such as depth-3 $\mathsf{AC}^0$ circuits, circuits with free gates, and nondeterministic branching programs (see \Cref{sec-related}); in particular, Nechiporuk's construction for nondeterministic branching programs~\cite{nechiporuk1962zero} translates directly into an encoding with $(4+o(1))\sqrt{2^n}$ clauses, which the upper bound of \Cref{prop:dense} improves.

\subsection{Upper bound for sparse functions}\label{sec:sparse-ub}
We now bootstrap \Cref{thm:sparse-simple} to prove the upper bound in \Cref{thm:sparse}.
Our construction is based on a succinct static dictionary, which
stores a function $f\colon\{\zero,\one\}^n\to\{\zero,\one\}$ with $|f^{-1}(\one)|=M$
using close to the information-theoretic minimum of $\lg\binom{2^n}{M}$ bits
and evaluates $f$ in $O(1)$ time on a word RAM.
Such dictionaries go back to Brodnik and Munro~\cite{BrodnikMunro1999}; Pagh~\cite{Pagh2001} and P{\u{a}}tra{\c{s}}cu~\cite{Patrascu2008} subsequently reduced the number of extra bits.
The following is the best known such dictionary, which we restate in a slightly weakened form convenient for our purposes.\footnote{Hu, Liang, Yu, Zhang, and Zhou~\cite[Theorem~1.1]{HuLYZZ2025} state their result as a static dictionary: in the word RAM model with word size $\Theta(\lg N)$, a set of $N$ keys from a universe of size $U\in[2N,\mathrm{poly}(N)]$ can be stored using $\lg\binom{U}{N}+\mathrm{polylog}(N)$ bits, given access to a fixed lookup table and to a hash function that can be described with $N^{\varepsilon}$ bits each, so that membership queries are answered in worst-case constant time. Their construction is randomized and succeeds with high probability, so in particular such a representation exists. We include the lookup table and the description of the hash function in $\sigma_f$, and apply this to the set $f^{-1}(\one)$ in the universe $\{\zero,\one\}^n$, whose size satisfies $2M\le2^n<M^2$.}
\begin{theorem}[Hu, Liang, Yu, Zhang, and Zhou~\cite{HuLYZZ2025}]\label{thm:static-dictionary}
    Let $f\colon\{\zero,\one\}^n\to\{\zero,\one\}$ with $|f^{-1}(\one)|=M$, where $2^{n/2}<M\le2^{n-1}$. There is a string $\sigma_f$ of
    \[
        \lg\binom{2^n}{M}+o(M)
    \]
    bits such that a word RAM with word size $\Theta(n)$ whose memory holds $\sigma_f$ computes $f(\vec{x})$ from $\vec{x}$ in worst-case $O(1)$ time.
\end{theorem}

Our encoding will be based on unrolling the lookup operations; since the lookup time is $O(1)$, the size of our encoding will be dominated by the cost of encoding the query algorithm's memory accesses (assuming $M$ is large). The key step is the following lemma, which may be of independent interest. If we identify a function with the binary string encoding its truth table, then the lemma shows that an entire contiguous block of bits can be fetched at essentially the same cost as fetching a single bit, even when the block is as long as the square root of the memory size.
Here, we only need blocks consisting of a single word of the RAM, i.e., $\Theta(n)$ bits.

\begin{lemma}\label{lem:sparse-word-table}
    Let $f\colon\{\zero,\one\}^n\to\{\zero,\one\}$, and let $w=2^k$ with $0\le k\le n$. For $j\in[0,2^{n-k}-1]$, the $j$th \emph{block} of $f$ is the string $\big(f(jw),f(jw+1),\ldots,f(jw+w-1)\big)\in\{\zero,\one\}^w$, where each argument of $f$ is identified with its $n$-bit binary representation. Define $g_w\colon\{\zero,\one\}^{n-k}\times\{\zero,\one\}^w\to\{\zero,\one\}$ by letting $g_w(j,\vec{z})=\one$ if and only if $\vec{z}$ is the $j$th block of $f$. Then,
    \[
        \enc(g_w)=O\left(\sqrt{2^n}+w\right).
    \]
\end{lemma}
\begin{proof}
    First suppose that $w\ge n$. Then $g_w$ has $n-k+w\le2w$ inputs. For each block index $j$, exactly one $\vec{z}$ satisfies $g_w(j,\vec{z})=\one$, so $|g_w^{-1}(\one)|=2^{n-k}$. By \Cref{thm:sparse-simple},
    \[
        \enc(g_w)=O\left(w+\sqrt{w\cdot2^{n-k}}\right)=O\left(\sqrt{2^n}+w\right).
    \]
    If $w<n$, let $w'$ be the power of two with $n\le w'<2n$. Then $\enc(g_w)\le\enc(g_{w'})+O(w')=O(\sqrt{2^n}+w)$.
\end{proof}

\noindent
We are now ready to prove the upper bound in \Cref{thm:sparse}.
\begin{lemma}\label{lem:sparse-ub}
    For all integers $n\ge1$ and $1\le M\le2^{n-1}$, we have $\enc(n,M)=O\left(n+\sqrt{M(n-\lg M)}\right)$.
\end{lemma}
\begin{proof}
    Let $f\colon\{\zero,\one\}^n\to\{\zero,\one\}$ with $|f^{-1}(\one)|=M$, where $1\le M\le2^{n-1}$. If $M\le2^{n/2}$, then $n-\lg M\ge n/2$, so \Cref{thm:sparse-simple} already gives $O(n+\sqrt{nM})=O(n+\sqrt{M(n-\lg M)})$ clauses. Hence, we may assume $M>2^{n/2}$. Since $2^n\le\binom{2^n}{M}\le(e2^n/M)^M$ and $n-\lg M\ge1$, the string $\sigma_f$ from \Cref{thm:static-dictionary} has $S=O(M(n-\lg M))$ bits, and note that $S \geq n$. Let the word size be a power of two $w=2^k=\Theta(n)=O(S)$, and let $f'\colon\{\zero,\one\}^{n'}\to\{\zero,\one\}$ be the function whose truth table is $\sigma_f$ padded with zeros to a length $2^{n'}=O(S)$ that is at least $w$. Then the words of memory are exactly the blocks of $f'$.

    The query algorithm runs a fixed program for $O(1)$ steps, so we can unroll it into a straight-line program with $O(1)$ memory reads and $O(1)$ word operations: at each step, perform every instruction of the program and keep the result of the one selected by the program counter. Truncating each read address to its $n'-k$ low-order bits keeps every read in range without affecting the actual computation. We introduce auxiliary variables for the register contents after each step, and take the conjunction of the encodings of the individual steps:
    \begin{itemize}
        \item each memory read is checked by the encoding of $g_w$ for $f'$ from \Cref{lem:sparse-word-table}, with $O(\sqrt{2^{n'}}+w)=O(\sqrt{M(n-\lg M)}+n)$ clauses;
        \item each word operation, including multiplication and division, is computed by a circuit with $O(n^2)$ gates, whose Tseitin encoding has $O(n^2)$ clauses.
    \end{itemize}
    Finally, a unit clause requires the output to be $\one$. Each step's encoding admits only the correct output for given inputs, so the resulting formula encodes $f$ with
    \[
        O\left(\sqrt{M(n-\lg M)}+n^2\right)
    \]
    clauses.
    Since $M>2^{n/2}$, we have $\sqrt{M(n-\lg M)}\ge2^{n/4}$, which asymptotically absorbs the $n^2$ term.
\end{proof}

\subsection{Lower bounds}\label{sec:arbitrary-lb}

We now prove the lower bounds in \Cref{thm:sparse} and \Cref{prop:dense}. Both are proved via an information-theoretic argument, using the counting bound of \Cref{lem-counting}.

\begin{lemma}\label{lem:sparse-lb}
    For all integers $n\ge1$ and $1\le M\le2^{n-1}$, we have
    \[
        \enc(n,M)\ge\max\left\{n,\frac12\sqrt{M(n-\lg M)}\right\} = \Omega\left(n+\sqrt{M(n-\lg M)}\right).
    \]
\end{lemma}
\begin{proof}
    Choose a support of size $M$ containing $\zero^n$ but none of the $n$ unit vectors. This is possible because $2^n-n\ge2^{n-1}$ for every $n\ge1$. Fix an accepting auxiliary assignment of any encoding at $\zero^n$. Flipping the $i$th input to obtain a unit vector must falsify some clause under the same auxiliary assignment. Such a clause had $\overline{x_i}$ as its only true literal at the original full assignment. One clause cannot have this property for two distinct input bits, so at least $n$ clauses are necessary. Consequently $\enc(n,M)\ge n$.

    Set $q:=\enc(n,M)\ge n\ge1$. By \Cref{lem-counting} and $\binom{2^n}{M}\ge(2^n/M)^M$,
    \[
        2^{M(n-\lg M)}\le\binom{2^n}{M}
        \le3^{(n+q)q}+1
        \le4^{(n+q)q}
        \le2^{4q^2}.
    \]
    Here we used $3^r+1\le4^r$ for positive integers $r$. Taking logarithms gives $q\ge\frac12\sqrt{M(n-\lg M)}$. 
\end{proof}

\begin{lemma}\label{lem:dense-lb}
    With $c := 1/\sqrt{\lg 3}$, we have $\enc(n)\ge(c-o(1))\sqrt{2^n}$.
\end{lemma}
\begin{proof}
    Let $q:=\enc(n)$. Every boolean function on $n$ inputs has an encoding with at most $q$ clauses, so \Cref{lem-counting} gives $2^{2^n}\le3^{(n+q)q}+1$. Solving for $q$, we obtain
    \[
        q\ge\sqrt{\frac{n^2}{4}+\log_3(2^{2^n}-1)}-\frac n2
        =c\sqrt{2^n}-\frac n2+o(1)
        =(c-o(1))\sqrt{2^n}. \qedhere
    \]
\end{proof}

\subsection{Upper bound for arbitrary functions}\label{sec:dense-ub}

We now give a constructive upper bound on $\enc(n)$, which is relatively similar to the construction of Dan\v{c}\'{i}k for depth-3 $\mathsf{AC}^0$ circuits~\cite{ac0}.
\begin{lemma}\label{lem:dense-ub}
    Every $f\colon \{\zero, \one\}^n \to \{\zero, \one\}$ satisfies $\enc(f)\le(2\sqrt{2}+o(1))\sqrt{2^n}$. In particular, $\enc(n)\le(2\sqrt{2}+o(1))\sqrt{2^n}$.
\end{lemma}
\begin{proof}
    Let $f\colon \{\zero, \one\}^n \to \{\zero, \one\}$ be arbitrary, and let $x_1, \ldots, x_n$ be input variables. First, introduce auxiliary variables $a_1, \dots, a_{\floor{n/2}}$ and $b_1, \dots, b_{\ceil{(n+1)/2}}$. Given an assignment \[ \tau \colon \{x_1,\dots,x_n,a_1,\dots,a_{\floor{n/2}},b_1, \dots, b_{\ceil{(n+1)/2}}\} \to \{\zero, \one\}, \] we can naturally define $X_\tau$, $A_\tau$, and $B_\tau$ to be the integers whose binary representations are given by the $x$-, $a$-, and $b$-variables respectively. Let $K := \floor{2^{(n+1)/2}}$. Then, we want to encode that any satisfying assignment $\tau$ will satisfy
    \[
        X_\tau = A_\tau \cdot K + B_\tau \qquad \text{and} \qquad B_\tau < K,
    \]
    which can be done by applying the Tseitin transformation to circuits for addition and multiplication, which results in $\mathrm{poly}(n)$ clauses.

    Let $\mathcal{A} := [0, \floor{(2^n-1)/K}]$. Every $X_\tau < 2^n$ satisfies $A_\tau = \floor{X_\tau/K} \in \mathcal{A}$. Moreover, $K \geq 2^{\ceil{n/2}}$, so every element of $\mathcal{A}$ is less than $2^{\floor{n/2}}$; hence distinct elements of $\mathcal{A}$ have distinct $\floor{n/2}$-bit binary representations. Next, for each $A \in \mathcal{A}$, introduce an auxiliary variable $y_A$. Let $A[i]$ be the $i$th bit of $A$'s binary representation. Then, add the following clauses:
    \begin{align}
     \forall i \in [\floor{n/2}],  \quad&   \left(\overline{a_i} \lor \bigvee_{\substack{A \in \mathcal{A}\\ A[i] = \one} } y_{A}\right) \land \left(a_i \lor \bigvee_{\substack{A \in \mathcal{A}\\ A[i] = \zero} } y_{A}\right)\label{eq:imp1}\\
        &\amo\lrp{\left\{y_A \mid A \in \mathcal{A}\right\}}.\label{eq:constraint-amos}
    \end{align}
Intuitively,~\eqref{eq:imp1} states that each $a_i$ variable implies a disjunction of the $A$-variables compatible with it, and then~\eqref{eq:constraint-amos} enforces that only one $A$-variable is assigned to $\one$.

Now, for each $B \in [0,K-1]$, let 
    \(
    S_{B} := \{ A \in \mathcal{A} \mid A \cdot K + B < 2^n \text{ and } f(A \cdot K + B) = \one  \},
    \)
    where by $f(A \cdot K + B)$ we mean the value of $f$ applied to the $n$-bit binary representation of $A \cdot K + B$. Intuitively, $S_B$ is the set of values of $A \in \mathcal{A}$ that are ``compatible'' with $B$.
    Moreover, for each $B \in [0,K-1]$, let 
    \[
    L_{B} := \{ b_i \mid B[i] = \one\} \cup \{ \overline{b_i} \mid B[i] = \zero\},
    \]
    where $B[i]$ is the $i$th bit of $B$'s binary representation. The encoding is then completed by adding clauses
    \begin{equation}\label{eq:cross-implications}
    \forall B \in [0,K-1], \quad\left(\bigvee_{A \in S_B}y_{A} \lor \bigvee_{\ell \in L_{B}} \overline{\ell} \right).
       \end{equation}
    
    First, let us analyze the number of clauses. The constraints enforcing that
    \[
        X_\tau = A_\tau \cdot K + B_\tau \qquad \text{and} \qquad B_\tau < K,
    \]
    for all $\tau$ use only $\mathrm{poly}(n) = o(\sqrt{2^n})$ clauses. Constraint~\eqref{eq:imp1} uses at most $n$ clauses, and by~\Cref{prop:amo-bound}, the constraint in~\eqref{eq:constraint-amos} can be encoded with $(2+o(1))|\mathcal{A}| = (\sqrt{2}+o(1))2^{n/2}$ clauses, since $|\mathcal{A}| \leq 2^n/K + 1$ and $K > 2^{(n+1)/2}-1$ give $|\mathcal{A}| = (1+o(1))2^{(n-1)/2}$. Finally,~\eqref{eq:cross-implications} includes $K = \floor{2^{(n+1)/2}} \le \sqrt{2} \cdot 2^{n/2}$ clauses. Thus, the total number of clauses is $(2\sqrt{2} + o(1))\sqrt{2^n}$.

    Now, to prove correctness, let $\tau\colon \{x_1, \dots, x_n\} \to \{\zero, \one\}$ be an arbitrary assignment, and let $X_\tau$ be the integer whose binary representation is given by the $x$-variables. Let $A_\tau$ and $B_\tau$ be the unique nonnegative integers such that
    \[
        X_\tau = A_\tau \cdot K + B_\tau \qquad \text{and} \qquad B_\tau < K.
    \]
    First, assume that $f(\tau(x_1), \dots, \tau(x_n)) = \one$. Let $Y := \{y_A \mid A \in \mathcal{A}\}$, and  consider the assignment $\theta\colon Y \to \{\zero, \one\}$ defined by
    \[
    \theta(y) = \begin{cases}
        \one & \text{if }y = y_{A_\tau}\\
        \zero & \text{otherwise}.
    \end{cases}
    \]
    Now, by correctness of the encoding for $\amo$, the assignment $\theta$ can be extended into an assignment $\theta'$ that assigns the auxiliary variables internal to~\eqref{eq:constraint-amos} satisfying said constraint. Then, extend $\theta'$ into an assignment $\theta''$ that assigns the $a$- and $b$-variables so that they are the binary representations of $A_\tau$ and $B_\tau$ respectively. Note that at this point the full assignment $\tau \cup \theta''$ satisfies~\eqref{eq:imp1} and~\eqref{eq:constraint-amos}. 
    Then, for each $B \in [0,K-1] \setminus \{B_\tau\}$ some literal in $L_B$ must be false, which satisfies~\eqref{eq:cross-implications}, and
    since $A_\tau \in S_{B_\tau}$,~\eqref{eq:cross-implications} is also satisfied for $B = B_\tau$.
    
    Conversely, assume $f(\tau(x_1), \dots, \tau(x_n)) = \zero$, and assume for a contradiction that some assignment $\theta$ of the auxiliary variables is such that $\tau \cup \theta$ satisfies all clauses. Then, the assignment to the $a$- and $b$-variables must be such that they are the binary representations of $A_\tau$ and $B_\tau$ respectively. Also, by correctness of the encoding of $\amo$ we have that $\theta(y_A) = \one$ for at most one $A \in \mathcal{A}$. Then, since $\tau \cup \theta$ satisfies~\eqref{eq:imp1} for $i = 1$ (any $i$ works), then by the resolution rule we have that at least one $y_{A}$ variable must be true. Thus, exactly one variable in $Y$ is assigned to $\one$ by $\theta$, and we claim that it must be $y_{A_\tau}$. Indeed, if it were $y_{A}$ for some $A \neq A_\tau$, and $i$ is an index such that $A[i] \neq A_\tau[i]$ (which exists because $A, A_\tau \in \mathcal{A}$), then~\eqref{eq:imp1} would not be satisfied for said $i$, giving us the desired contradiction.
 Now, if the only true variable in $Y$ is $y_{A_\tau}$, then constraint~\eqref{eq:cross-implications} is not satisfied for $B = B_\tau$, since by assumption $A_\tau \not\in S_{B_\tau}$, which yields the desired contradiction.
\end{proof}

\section{A Characterization of Encoding Complexity}\label{sec-characterization}
A \emph{De Morgan circuit} is a circuit over the basis $\{\land,\lor,\lnot\}$ such that $\lnot$ is only applied to input variables. We allow constants $\zero$ and $\one$, and we allow both $\land$-gates and $\lor$-gates to have unbounded fan-in. In other words, a De Morgan circuit on the input variables $x_1, \dots, x_n$ is a monotone circuit over the basis $\{\land,\lor\}$ on the input \emph{literals} $x_1, \dots, x_n, \overline{x_1}, \dots, \overline{x_n}$. If a De Morgan circuit $C$ on the input variables $x_1, \dots, x_n, y_1, \dots, y_m$ computes a function $g \colon \{\zero,\one\}^{n+m} \to \{\zero, \one\}$ such that
\[
\forall \vec{x} \in \{\zero,\one\}^n, \quad f(\vec{x}) = \one \Longleftrightarrow \exists{\vec{y} \in \{\zero,\one\}^m}.\; g(\vec{x},\vec{y}) = \one,
\]then we say that $C$ is a \emph{nondeterministic De Morgan circuit} computing $f$. Our metric for these circuits, which we will show coincides with CNF encoding complexity, is the number of wires entering its $\land$-gates:
\begin{definition}
    Given a boolean function $f \colon \{\zero,\one\}^n \to \{\zero, \one\}$, let $\textsf{nondet-wires}_{\land}(f)$ be the minimum, over all nondeterministic De Morgan circuits computing $f$, of the total fan-in of their $\land$-gates.
\end{definition}

We prove in \Cref{thm-enc-characterization} that this measure coincides exactly with encoding complexity, except for functions with $\enc(f) = 1$, which are computed by a single $\lor$-gate and hence need no $\land$-gates.

\begin{proposition} \label{thm-enc-characterization}
    For every boolean function $f$,
    \[
        \textsf{nondet-wires}_{\land}(f)=
        \begin{cases}
            \enc(f) & \text{if } \enc(f)\ge2,\\
            0 & \text{if } \enc(f)\le1.
        \end{cases}
    \]
\end{proposition}
\begin{proof}
    Write $W:=\textsf{nondet-wires}_{\land}(f)$. First, we show that $W\le\enc(f)$ if $\enc(f)\ge2$, and $W=0$ if $\enc(f)\le1$. Take an encoding of $f$ with $c=\enc(f)$ clauses, and regard its auxiliary variables as nondeterministic inputs. If $c\ge2$, regard each clause as an $\lor$-gate and combine them with a single $\land$-gate of fan-in $c$. If $c=1$, the single clause, regarded as an $\lor$-gate, is the whole circuit. If $c=0$, then $f$ is constantly $\one$, and the constant $\one$ is the whole circuit.

    It remains to show that every nondeterministic De Morgan circuit $C$ computing $f$ whose $\land$-gates have total fan-in $w$ satisfies $\enc(f)\le\max\{1,w\}$. If $f$ is constantly $\one$, then $\enc(f)=0$, so assume otherwise. Remove all gates that do not reach the output; this does not increase $w$. Write $o$ for the output gate of $C$. For every gate or input $u$ of $C$, define its \emph{disjunct set} $D(u)$ as follows: if $u$ is a literal, a constant, or an $\land$-gate, then $D(u):=\{u\}$; if $u$ is an $\lor$-gate, then $D(u)$ is the union of the disjunct sets of its inputs. Thus, every $\lor$-gate computes the disjunction of its disjunct set. For a disjunct set $S$ not containing $\one$, let $\mathrm{cl}(S)$ be the clause obtained from $S$ by replacing every $\land$-gate $h$ by a variable $z_h$, keeping every literal, and deleting $\zero$ if it is present. Since $f$ is not constantly $\one$, the disjunct set $D(o)$ does not contain $\one$.

    If $C$ has no $\land$-gates, then $D(o)$ consists of literals and possibly $\zero$, and $C$ outputs $\one$ exactly when $\mathrm{cl}(D(o))$ is true. Hence the single clause $\mathrm{cl}(D(o))$, with the nondeterministic inputs of $C$ as auxiliary variables, is an encoding of $f$, and $\enc(f)\le1$.

    Otherwise, fix an $\land$-gate $g^\star$ such that every path from $g^\star$ to $o$ passes through no other $\land$-gates. Then, $g^\star\in D(o)$, and $g^\star$ does not lie in any disjunct set of an input of an $\land$-gate. Let $B:=\mathrm{cl}(D(o)\setminus\{g^\star\})$, and introduce a fresh auxiliary variable $z_g$ for every $\land$-gate $g\neq g^\star$. Our encoding consists of the clauses
    \begin{align*}
        &\overline{z_g}\lor\mathrm{cl}(D(v)) &&\text{for every $\land$-gate $g\neq g^\star$ and every input $v$ of $g$ with $\one\notin D(v)$},\\
        &B\lor\mathrm{cl}(D(v)) &&\text{for every input $v$ of $g^\star$ with $\one\notin D(v)$},
    \end{align*}
    and its auxiliary variables are the nondeterministic inputs of $C$ and the variables $z_g$ for every $\land$-gate $g \neq g^\star$. There is at most one clause per wire entering an $\land$-gate, so the encoding has at most $w$ clauses.

    Now we show that the encoding is correct. First, suppose $\vec{x}$ is such that $f(\vec{x})=\one$, and fix nondeterministic inputs for which $C$ outputs $\one$, and set each $z_g$ to the value of $g$. If an $\land$-gate evaluates to $\one$, then so does each of its inputs $v$, so some element of $D(v)$ is true, and this element makes $\mathrm{cl}(D(v))$ true. This satisfies the clauses of every $\land$-gate $g\neq g^\star$, and also the clauses of $g^\star$ if $g^\star$ evaluates to $\one$. Otherwise, since $o$ evaluates to $\one$, some element of $D(o)\setminus\{g^\star\}$ is true, which makes $B$ true. Thus, every clause is satisfied.

    Conversely, suppose the encoding is satisfiable for an input $\vec{x}$, and fix a satisfying assignment to the auxiliary variables. We show that $C$ evaluates to $\one$ on input $\vec{x}$ when the nondeterministic inputs are set according to the satisfying assignment; this implies that $f(\vec{x})=\one$.

    We first prove the following claim: for every $\land$-gate $g\neq g^\star$, if $z_g=\one$, then $g$ evaluates to $\one$. We proceed by induction along a topological order of $C$. Let $g\neq g^\star$ be an $\land$-gate with $z_g=\one$, and assume the claim for all $\land$-gates that precede $g$. Let $v$ be an input of $g$; we show that $v$ evaluates to $\one$. If $\one\in D(v)$, this holds because $v$ computes the disjunction of $D(v)$. Otherwise, the encoding contains the clause $\overline{z_g}\lor\mathrm{cl}(D(v))$, and since $z_g=\one$, some literal of $\mathrm{cl}(D(v))$ is true. Either this literal is an element of $D(v)$, or it is $z_h$ for an $\land$-gate $h\in D(v)$. In the latter case, $h$ precedes $g$ and $h\neq g^\star$, so $h$ evaluates to $\one$ by the induction hypothesis. Thus, some element of $D(v)$ evaluates to $\one$, and therefore so does $v$. Since every input of $g$ evaluates to $\one$, so does $g$, which proves the claim.

    Consequently, if $\mathrm{cl}(S)$ is true, where $S$ is a set of literals, $\land$-gates other than $g^\star$, and possibly $\zero$, then some element of $S$ evaluates to $\one$. We use this to show that $o$ evaluates to $\one$, distinguishing two cases. If $B$ is true, then some element of $D(o)\setminus\{g^\star\}$ evaluates to $\one$, so $o$ evaluates to $\one$. If $B$ is false, then for every input $v$ of $g^\star$ with $\one\notin D(v)$, the clause $B\lor\mathrm{cl}(D(v))$ forces $\mathrm{cl}(D(v))$ to be true. Since $g^\star\notin D(v)$, some element of $D(v)$ evaluates to $\one$, and hence so does $v$; inputs $v$ with $\one\in D(v)$ evaluate to $\one$ trivially. Thus, every input of $g^\star$ evaluates to $\one$, so $g^\star$ evaluates to $\one$. Since $g^\star\in D(o)$, $o$ evaluates to $\one$. In both cases, $C$ outputs $\one$. Therefore, the encoding is correct.
\end{proof}

\section*{Acknowledgments}

We thank Ryan O'Donnell for helpful conversations about the presentation of our results.
Krapivin was supported in part by NSF grant 2504471, NSF CAREER award 2542165, a Jane Street Research Grant, the Jeanne~B.\ and Richard~F.\ Berdik ARCS Pittsburgh Endowed Scholar Award, the David and Lucile Packard Foundation (Award \#202071730), and gifts from Nikolai Mushegian and Microsoft.
Przybocki and Subercaseaux were supported by the DARPA expMath program through the DARPA CMO contract number HR0011262E028 and by the National Science Foundation (NSF) under grant
DMS2434625. Przybocki was additionally supported by the NSF Graduate Research Fellowship Program under Grant Nos. DGE-2140739 and DGE-2631988. 

\section*{AI Methodology}

We used ChatGPT 6 Astra in order to make quantitative improvements on some of our results. Concretely, we provided Astra with a human-written proof of a version of~\Cref{thm-polytime} that used Szemer\'edi's regularity lemma and yielded $O_k\lrp{\frac{n^k \cdot \lg(\lg^*{n})}{\lg(n) \sqrt[5]{\lg^*(n)}}}$ clauses for arbitrary $k$-CNF. Astra provided~\Cref{lem-sparse-rectangle-search} and \Cref{lem-decomp} in order to yield the improved bound presented in the paper. Similarly, we presented it with a weaker version of~\Cref{thm:sparse}, which gave an upper bound of $O(\sqrt{M}\cdot(n - \lg M + 1))$, and Astra improved this by leveraging a succinct static dictionary.
In both cases, the original writing produced by the model was poor, and some of its arguments were overcomplicated, so we simplified and rewrote each of them. We also used Claude Opus 5.5 for help with writing and reorganizing some proofs.

\bibliographystyle{alphaurl}
\bibliography{bib}

\end{document}